\documentclass[11pt,a4paper]{article}

\pdfoutput=1
\usepackage{jcappub}
\usepackage{graphicx}

\usepackage{amssymb}
\usepackage{amsmath}
\usepackage{bm}
\usepackage{latexsym}
\usepackage{epsfig}
\usepackage{psfrag}
\usepackage[normalem]{ulem}
\usepackage{textcomp}
\usepackage{color}
\usepackage{pstricks}
\usepackage[utf8]{inputenc}
\allowdisplaybreaks

\makeatletter
\gdef\@fpheader{}
\makeatother

\def\nn{\nonumber} 

\def\f{\frac}
\def\l{\left}
\def\r{\right}
\def\d{{\rm d}}
\def\Mpl{M_{_{\mathrm{Pl}}}}
\def\Mp{M_{_{\mathrm{Pl}}}}

\def\HI{H_{_{\mathrm I}}}

\def\ps{{\mathcal P}_{_{\mathrm{S}}}}

\def\ns{n_{_{\mathrm{S}}}}
\def\pc{{\mathcal P}_{_{\mathrm{C}}}}
\def\ph{{\mathcal P}_{h}}

\def\kf{k_{\mathrm f}}
\def\kT{k_{_{\mathrm{T}}}}

\def\cR{{\mathcal R}}

\def\ei{\eta_{\rm i}}

\def\ee{\eta_{\rm e}}

\def\fpbh{f_{_{\mathrm{PBH}}}}
\def\ogw{\Omega_{_{\mathrm{GW}}}}
\def\fnl{f_{_{\rm NL}}}

\def\vka{{\bm k}_{1}}
\def\vkb{{\bm k}_{2}}
\def\vkc{{\bm k}_{3}}

\def\vx{\textbm{x}}
\def\vk{\textbm{k}}
\def\vp{\bm{p}}

\def\vx{{\bm{x}}}
\def\vk{{\bm{k}}}

\def\mpcinv{{\mathrm{Mpc}}^{-1}}
\def\cI{{\mathcal I}}

\begin{document}
\title{Could PBHs and secondary GWs have originated from squeezed initial states?}
\author[a]{H.~V.~Ragavendra,}
\author[a]{L.~Sriramkumar,}
\author[b,c,d,e]{and Joseph Silk}
\affiliation[a]{Department of Physics, Indian Institute of Technology Madras, 
Chennai~600036, India}
\affiliation[b]{Institut d'Astrophysique de Paris, UMR 7095, CNRS/UPMC Universit\'e 
Paris~6, Sorbonne Universit\'es, 98 bis boulevard Arago, F-75014 Paris, France}
\affiliation[c]{Institut Lagrange de Paris, Sorbonne Universit\'es, 
98 bis Boulevard Arago, 75014 Paris, France}
\affiliation[d]{ Department of Physics and Astronomy, The Johns Hopkins University, 
3400 N.~Charles Street, Baltimore, MD 21218, U.S.A.}
\affiliation[e]{Beecroft Institute for Cosmology and Particle Astrophysics,
University of Oxford, Keble Road, Oxford OX1 3RH, U.K.}
\emailAdd{ragavendra@physics.iitm.ac.in}
\emailAdd{sriram@physics.iitm.ac.in}
\emailAdd{silk@iap.fr}

\abstract{Recently, the production of primordial black holes (PBHs) and 
secondary gravitational waves (GWs) due to enhanced scalar power on small 
scales have garnered considerable attention in the literature. 
Often, the mechanism considered to arrive at such increased power involves 
a modification of the standard slow roll inflationary dynamics, achieved 
with the aid of fine-tuned potentials.
In this work, we investigate another well known method to generate features 
in the power spectrum wherein the initial state of the perturbations is 
assumed to be squeezed states. 
The approach allows one to generate features even in slow 
roll inflation with a specific choice for the Bogoliubov coefficients 
characterizing the squeezed initial states.
Also, the method is technically straightforward to implement since the 
Bogoliubov coefficients can be immediately determined from the form 
of the desired spectrum with increased scalar power at small scales.
It is known that, for squeezed initial states, the scalar bispectrum is 
strongly scale dependent and the consistency condition governing the scalar 
bispectrum in the squeezed limit is violated.
In fact, the non-Gaussianity parameter $\fnl$ characterizing the scalar 
bispectrum proves to be inversely proportional to the squeezed mode and 
this dependence enhances its amplitude at large wave numbers making it 
highly sensitive to even a small deviation from the standard Bunch-Davies 
vacuum.
These aspects can possibly aid in leading to enhanced formation of PBHs 
and generation of secondary GWs.
{\it However, we find that: (i)~the desired form of the
squeezed initial states may be challenging to achieve from a dynamical 
mechanism, and (ii)~the backreaction due to the excited states severely 
limits the extent of deviation from the Bunch-Davies vacuum at large wave
numbers.}\/
We argue that, unless the issue of backreaction is circumvented,
squeezed initial states {\it cannot}\/ lead to a substantial increase in power 
on small scales that is required for enhanced formation of PBHs and generation 
of secondary GWs.}

\maketitle


\section{Introduction}

It is now almost half-a-century since it was originally argued that 
black holes could have formed due to over-densities in the primordial 
universe~\cite{Carr:1974nx,Carr:1975qj}.
The investigations of such primordial black holes (PBHs) have gained 
traction over the last few years with the observations of gravitational 
waves (GWs) from the mergers of binary black 
holes~\cite{Abbott:2016nmj,Abbott:2017vtc,Abbott:2017gyy,Abbott:2020tfl}. 
Several current and upcoming observational efforts promise to provide 
constraints on the fraction of the PBHs constituting the bulk of cold
dark matter density in the current universe, a quantity usually referred 
to as~$\fpbh$~\cite{Carr:2020gox}.
Motivated by these observational efforts, there has been several attempts 
to build models of inflation that could generate considerable population 
of PBHs over certain mass ranges (see, for example, refs.~\cite{Domcke:2017fix,
Garcia-Bellido:2017mdw,Ballesteros:2017fsr,Germani:2017bcs,Dalianis:2018frf}).

It is well known that scales smaller than those associated with the cosmic 
microwave background (CMB), say, with wave numbers $k>1\,\mpcinv$, 
reenter the Hubble radius during the radiation dominated epoch. 
If the scalar power over these small scales have enhanced amplitudes (when
compared to their COBE normalized values over the CMB scales), they could, 
in principle, induce instantaneous collapses of energy densities of 
corresponding sizes, thereby forming PBHs~\cite{Chongchitnan:2006wx,
PinaAvelino:2005rm,Inomata:2017okj}. 
To achieve a higher amplitude in the inflationary scalar perturbation 
spectrum (say, of the order of~$10^{-2}$) at larger wave numbers, one 
has to suitably model the background dynamics so that a departure from 
slow roll inflation arises at late times. 
It has been found that, in single field models, inflationary potentials 
containing a point of inflection can generate the required boost in the 
scalar power (see, for instance, refs.~\cite{Germani:2017bcs,Bhaumik:2019tvl,
Ragavendra:2020sop,Bhaumik:2020dor}). 
The inflection point in the potential leads to a transient epoch of ultra 
slow roll inflation, which turns out to be responsible for the rise in the 
scalar power over small scales. 
Other features, such as a bump or dip artificially added to the potential 
are also known to boost the scalar power at larger wave 
numbers~\cite{Atal:2019cdz,Mishra:2019pzq}. 
There have also been attempts to generate PBHs using other mechanisms such
as models involving non-canonical scalar
fields~\cite{Kamenshchik:2018sig,Ballesteros:2018wlw}, 
inflation driven by multiple 
fields~\cite{Palma:2020ejf,Fumagalli:2020adf,Braglia:2020eai,Zhou:2020kkf,
Fumagalli:2020nvq}, inducing a non-trivial speed of sound during
inflation~\cite{Cai:2018tuh,Chen:2019zza,Romano:2020gtn}, or
a modified history of reheating and radiation dominated era following
inflation~\cite{Carr:2018nkm,Bhattacharya:2019bvk}.

Moreover, when the scalar power is boosted to large amplitudes, the second order 
tensor perturbations that are sourced by the quadratic terms involving the first 
order scalar perturbations can dominate the contributions due to the original,
inflationary, first order tensor perturbations~\cite{Baumann:2007zm,Ananda:2006af}. 
In other words, the enhanced scalar power, apart from producing a significant 
amount of PBHs, also leads to considerable amplification of the secondary GWs 
at small scales or, equivalently, at large frequencies~\cite{Clesse:2018ogk}.
These GWs induced by the scalar perturbations are expected to be stochastic 
and isotropic. 
There are several experiments and observational surveys that constrain the 
dimensionless energy density of such a stochastic gravitational wave background, 
say, $\ogw$, observable today~\cite{Moore:2014lga}.

As we mentioned above, the enhancement in the scalar power over small scales 
can be achieved with the aid of a brief period of departure from slow roll 
inflation.
We should point out here that such scenarios would also produce a strongly 
scale dependent bispectrum. 
However, it has been shown that, in single field models of inflation wherein 
the deviation from slow roll is brief, the consistency condition relating the
bispectrum and the power spectrum in the squeezed limit is indeed satisfied
(in this context, see refs.~\cite{Sreenath:2014nca,Ragavendra:2020sop,
Bravo:2020hde}). 
This implies that the magnitude of the scalar non-Gaussianity parameter, 
$\fnl$, is at the most of order unity over the range of wave numbers which 
contains enhanced power. 
As a result, any corrections due to the bispectrum that has to be 
accounted for in the power spectrum proves to be negligible in these 
models~\cite{Ragavendra:2020sop}.

However, the aforementioned methods of modifying slow roll inflation to 
achieve sufficient enhancement in the scalar power, and hence produce 
significant amount of PBHs and secondary GWs, are known to pose certain 
challenges.
They typically require extreme fine-tuning of the parameters involved.
Else, they may either prolong the duration of inflation beyond reasonable 
number of e-folds or alter the scalar spectral index~$\ns$ and the 
tensor-to-scalar ratio~$r$ over the CMB scales thereby leading to a 
tension with the constraints from Planck data (see, for instance,
refs.~\cite{Germani:2017bcs,Ragavendra:2020sop}).
There exists another approach to achieve power spectra with 
the desired shape at small scales. 
The alternative method is to work with non-vacuum, specifically, squeezed, 
initial states for the perturbations during inflation.
This method of evolving the perturbations with initial states other than 
the standard Bunch-Davies vacuum is well known in the literature and has 
been discussed in various contexts (see, for example,
refs.~\cite{Brandenberger:2002hs,Sriramkumar:2004pj,Holman:2007na,Meerburg:2009ys,
Meerburg:2009fi,Agullo:2010ws,Meerburg:2011gd,Ganc:2011dy,Kundu:2011sg,
Brandenberger:2012aj,Kundu:2013gha,Shukla:2016bnu,Seleim:2020eij}). 
These excited initial states for the perturbations can be expressed in terms 
of the so-called Bogoliubov coefficients. 
As we shall see, the Bogoliubov coefficients essentially provide us an 
independent function to introduce the desired features in the power spectrum.
However, while it is technically straightforward to arrive at the
required power spectrum with a suitable choice of the Bogoliubov coefficients,
we encounter two drawbacks with the proposed approach.
On the one hand, it seems challenging to design a mechanism that leaves the 
curvature perturbations in such an excited initial state.
On the other hand, we find that squeezed initial states lead to significant 
backreaction during the early stages of inflation unless the state is 
remarkably close to the Bunch-Davies vacuum.

To illustrate these points, in this work, we shall focus 
on the popular lognormal shape of amplification in the scalar power 
spectrum~\cite{Clesse:2018ogk,Pi:2020otn}. 
In the following section, we shall briefly describe the modes corresponding to
squeezed initial states and discuss the corresponding scalar power and bispectra. 
We shall consider suitable functional forms for the Bogoliubov coefficients
to produce the lognormal feature in the power spectrum and calculate the 
corresponding scalar bispectrum analytically. 
We shall show that the bispectrum is significantly enhanced in the squeezed 
limit and that the consistency condition is strongly violated over the range 
of wave numbers containing the lognormal feature.
In other words, we find that the cubic order non-Gaussian modifications to 
the scalar power spectrum can possibly dominate the amplitude of the
original scalar power around the feature for certain values of the parameter 
that characterizes the deviations from the Bunch-Davies vacuum.
In section~\ref{sec:o}, we shall compute the observable quantities of 
interest, viz. $\fpbh$ and $\ogw$, generated from such an enhanced scalar
power spectrum. 
In section~\ref{sec:c}, we shall first discuss possible mechanisms that can 
lead to the squeezed initial states for the curvature perturbation at early 
times.
Thereafter, we shall describe the issue of backreaction wherein we compute the 
energy density associated with the perturbations evolved from squeezed initial
states and compare it against the background energy density.
We argue that it is rather challenging to achieve such specific initial states
by invoking mechanisms operating prior to inflation.
Moreover, we find that the backreaction severely restricts the extent of deviation 
of the initial state from the Bunch-Davies vacuum, particularly on small scales. 
This, in turn, implies that the desired amplification in the power spectrum 
and the larger levels of non-Gaussianities {\it cannot}\/ be achieved in this 
approach unless the choice of the specific initial state is satisfactorily
justified and the issue of backreaction is overcome.
We shall finally conclude in section~\ref{sec:f} with a brief summary and 
outlook.

Before we proceed, we should clarify the conventions and notations that we 
shall adopt in this work. 
We shall work with natural units such that $\hbar=c=1$ and set the reduced 
Planck mass to be $\Mpl=\l(8\,\pi\, G\r)^{-1/2}$.
We shall assume the background to be the spatially flat 
Friedmann-Lema\^itre-Robertson-Walker~(FLRW) line element described by 
the scale factor~$a$ and the Hubble parameter~$H$.
Note that $\eta$ shall represent the conformal time coordinate and an overprime 
shall denote differentiation with respect to~$\eta$.


\section{Squeezed initial states, scalar power and bispectra}

In this section, we shall construct scalar power spectra with a lognormal peak
from squeezed initial states.
We shall also calculate the associated scalar bispectra and utilize the result 
to arrive at the corresponding non-Gaussian modifications to the power spectrum.

As far as the background dynamics is concerned, we shall have in mind the scenario
of slow roll inflation. 
Recall that, in such a case, while it is the combination of the nearly constant 
Hubble parameter~$\HI$ and the first slow parameter~$\epsilon_1$ that
determine the amplitude of the scalar power spectrum, the first two slow roll 
parameters~$\epsilon_1$ and~$\epsilon_2$ determine the scalar spectral index~$\ns$.
Moreover, the tensor-to-scalar ratio~$r$ is determined by the first slow roll 
parameter~$\epsilon_1$.
The values of these parameters can be chosen so that we achieve nearly scale 
invariant scalar and tensor power spectra that are consistent with the recent 
constraints from Planck over the CMB scales~\cite{Akrami:2018odb}.
However, for convenience, in our calculations below, we shall work with the de 
Sitter modes to describe the scalar perturbations.
The modes, say, $f_k(\eta)$, describing the scalar perturbations that emerge 
from initial conditions corresponding to squeezed states can be expressed 
as~\cite{Brandenberger:2002hs,Sriramkumar:2004pj,Holman:2007na,Meerburg:2009ys,
Meerburg:2009fi,Agullo:2010ws,Meerburg:2011gd,Ganc:2011dy,Kundu:2011sg,
Brandenberger:2012aj,Kundu:2013gha,Seleim:2020eij} 
\begin{equation}
f_k(\eta) 
= \frac{i\,\HI}{2\,\Mp\,\sqrt{k^3\,\epsilon_1}}\,
\l[\alpha(k)\,\l(1+i\,k\,\eta\r)\, \mathrm{e}^{-i\,k\,\eta}
-\beta(k)\,\l(1-i\,k\,\eta\r)\,\mathrm{e}^{i\,k\,\eta}\r],
\label{eq:ds-m}
\end{equation}
where $\alpha(k)$ and $\beta(k)$ are the so-called Bogoliubov coefficients. 
Note that the standard Bunch-Davies initial conditions correspond to 
setting $\alpha(k)= 1$ and $\beta(k) = 0$.
The above modes correspond to squeezed initial states that are excited states 
above the Bunch-Davies vacuum.  
We should also mention that the Bogoliubov coefficients $\alpha(k)$ and $\beta(k)$ 
are not completely independent functions, but satisfy the following constraint:
\begin{equation}
\vert\alpha(k)\vert^2 - \vert\beta(k)\vert^2 = 1.\label{eq:ab-cond}
\end{equation}
This constraint arises due to the fact that the Wronskian associated with the 
differential equation governing the scalar perturbations is a constant, which
is determined by the initial conditions imposed on the modes. 


\subsection{Power spectrum from squeezed initial states}

The power spectrum of the scalar perturbations evolving from squeezed initial
states can be evaluated towards the end of inflation (i.e. as $\eta\to 0$). 
Upon using the modes~\eqref{eq:ds-m}, the resulting power spectrum can be 
expressed in terms of the Bogoliubov coefficients~$\alpha(k)$ and~$\beta(k)$ 
as follows:
\begin{equation}
\ps(k) 
= \f{k^3}{2\,\pi^2}\,\vert f_k(\eta \to 0)\vert^2
=\ps^0(k)\,\vert\alpha(k) - \beta(k)\vert^2,\label{eq:ps-ab}
\end{equation}
where 
\begin{equation} 
\ps^0(k) = \f{\HI^2}{8\,\pi^2\,\Mpl^2\,\epsilon_1}
\end{equation}
is the COBE normalized, nearly scale invariant spectrum with a small red tilt. 
Since we are interested in the small scale features of the spectrum, for simplicity,
we shall assume that $\ps^0(k)$ is strictly scale invariant with a COBE normalized 
amplitude over all the wave numbers of our interest. 
We should hasten to add that introducing a small red tilt does not affect our 
conclusions in the remainder of our discussion.
We shall choose to work with the following values of the primary slow roll
inflationary parameters:~$\HI=4.16\times10^{-5}\,\Mpl$, $\epsilon_1=10^{-2}$ 
and $\epsilon_2=2\,\epsilon_1$.
Also, note that the power spectrum is independent of an overall phase factor and
depends only on the relative phase factor between $\alpha(k)$ and $\beta(k)$. 

Let us now define $\delta(k)=\beta(k)/\alpha(k)$.
Then, upon using the constraint~\eqref{eq:ab-cond}, the power spectrum~\eqref{eq:ps-ab} 
can be written in terms of the function~$\delta(k)$ as
\begin{equation}
\ps(k) = \ps^0(k)\,
\l[\frac{\vert 1-\delta(k)\vert^2}{1-\vert\delta(k)\vert^2}\r].
\end{equation}
For ease of modeling, we shall assume the relative phase factor between~$\alpha(k)$ 
and $\beta(k)$ to be zero.
We should clarify that this assumption is made just to simplify our calculations.
It can be relaxed, if needed, to model the spectrum with the phase factor taken 
into account. 
Setting the relative phase factor to be zero essentially implies that $\delta(k)$ 
is real so that the above expression for the scalar power spectrum reduces to 
\begin{equation}
\ps(k) = \ps^0(k)\,\l\{\frac{\l[1-\delta(k)\r]^2}{1-\delta^2(k)}\r\}.\label{eq:ps-f}
\end{equation}

With the above form of the spectrum arising from squeezed initial states, we shall
now proceed to model the feature of our interest. 
Let us assume that the power spectrum has a localized feature over a certain range 
of wave numbers, say, $g(k)$, so that~$\ps(k)$ is given by
\begin{equation}
\ps(k) = \ps^0(k)\,\l[1+g(k)\r].\label{eq:gk}
\end{equation}
Upon comparing the above two equations, it is evident that the feature $g(k)$ is 
related to~$\delta(k)$ as follows:
\begin{equation}
\delta(k) = \,\frac{-g(k)}{2+g(k)}\label{eq:deltak}
\end{equation}
It should be clear that we have essentially traded off the function~$g(k)$ 
for~$\delta(k)$.
In other words, we can choose an initial squeezed state described by~$\delta(k)$ 
to lead to the desired feature~$g(k)$ in the power spectrum.
In this work, we shall assume $g(k)$ to be a lognormal function of the wave
number~$k$. 
Such a form for the feature in the spectrum is often considered because of the 
fact that, when departures from slow roll arise,  many single field and two field 
models lead to  scalar power spectra whose shape near the peak can be roughly 
approximated by such a function (see, for instance,
refs.~\cite{Clesse:2015wea,Pi:2020otn,Braglia:2020eai}). 
Also, it simplifies the calculations involved and hence allows an easier 
comparison of the quantities~$\fpbh$ and $\ogw$ against the observational
constraints~\cite{Pi:2020otn,Gow:2020cou}. 
We shall assume that the function $g(k)$ takes the form
\begin{equation}
g(k) = \f{\gamma}{\sqrt{2\,\pi\, \Delta_k^2}}\,
{\mathrm{exp}}\l[-\displaystyle\frac{{\rm 
ln^2}(k/\kf)}{2\,\Delta^2_k}\r],
\label{eq:gk-ln}
\end{equation}
where $\gamma$ represents the strength of the feature in the spectrum, $\Delta_k$ 
determines the width of the Gaussian and $\kf$ denotes the location of the peak of 
the lognormal distribution.
It is useful to note here that, given $g(k)$, the Bogoliubov coefficients~$\alpha(k)$ 
and~$\beta(k)$ can be obtained to be
\begin{equation}
\alpha(k) = \frac{2 + g(k)}{2\,\sqrt{1+g(k)}},\quad
\beta(k) = \frac{-g(k)}{2\,\sqrt{1+g(k)}}.\label{eq:al-be}
\end{equation}
We should stress again that these expressions for~$\alpha(k)$ and~$\beta(k)$ 
have been arrived at under the assumption that their relative phase factor is
zero.
We should also point out that setting $\gamma = 0$ leads to $g(k) = 0$, $\delta(k)=0$, 
$\alpha(k) = 1$ and $\beta(k)=0$.
This recovers the standard Bunch-Davies vacuum state and the scale invariant spectrum. 
Moreover, note that, for modes far away from $\kf$, i.e. for $k\gg \kf$ or $k\ll \kf$,
$g(k)\rightarrow 0$, and we again recover the standard Bunch-Davies vacuum state. 
Therefore, it should be clear that, in our scenario, it is only modes around 
$k_\mathrm{f}$ which evolve from non-vacuum initial states.
Further, the strength of their deviation from the vacuum state is proportional to
the parameter $\gamma$.

\begin{figure}[!t]
\centering
\includegraphics[width=15cm]{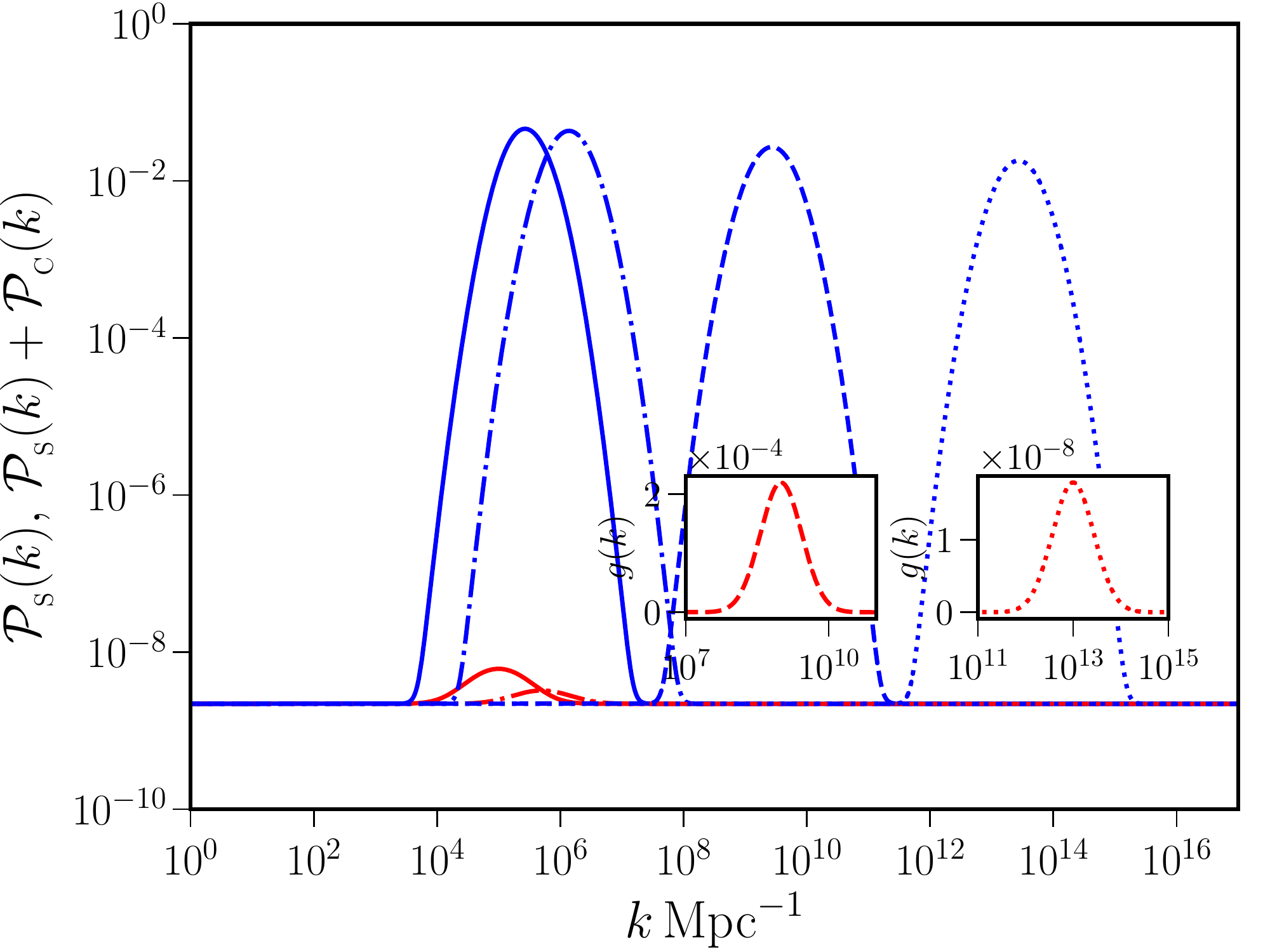}
\caption{The scalar power spectra with a lognormal shape obtained from suitably 
chosen squeezed initial states have been plotted for different sets of the 
parameters~$\gamma$ and~$\kf$ that determine the strength and the location of
the peaks.
Note that, we have plotted the original spectra $\ps(k)$ (in red) as well 
as the modified spectra $\ps(k)+\pc(k)$ (in blue), where $\pc(k)$ denotes
the non-Gaussian modifications to the power spectrum [cf. eqs.~\eqref{eq:psc} 
and \eqref{eq:pc-k}].
We have illustrated the spectra for the following four values
of~$\kf$:~$10^5\,\mpcinv$ (as solid curves), $5\times10^5\,\mpcinv$ (as 
dashed-dotted curves), $10^9\,\mpcinv$ (as dashed curves) and $10^{13}\,
\mpcinv$ (as dotted curves).
We have chosen the corresponding values of~$\gamma$ to be $4.5$, $1.2$, 
$5.5\times10^{-4}$ and $4.5\times10^{-8}$, respectively. 
We have set the width $\Delta_k$ of the lognormal distribution to be unity in 
all the cases.
The features in the original spectra $\ps(k)$~with peaks around $10^9\,\mpcinv$ 
and $10^{13}\, \mpcinv$ are not as discernible as those at the two other locations 
due to the small values of~$\gamma$.
Hence, in these two cases, we have included insets to highlight 
the function $g(k)$ [cf. eq.~\eqref{eq:gk-ln}] instead.
The parameter~$\gamma$ has been chosen so that, when the non-Gaussian modifications
are taken into account, all the power spectra have roughly the same amplitudes at 
their peaks.}\label{fig:ps}
\end{figure}

In figure~\ref{fig:ps}, we have plotted the scalar power spectra~$\ps(k)$ 
containing a lognormal feature with peaks located at four different wave 
numbers~$\kf$ with suitable values for the parameter~$\gamma$. 
In the figure, we have also plotted the modified power spectra, 
i.e.~$\ps(k)+\pc(k)$ [cf. eqs.~\eqref{eq:psc} and \eqref{eq:pc-k}], 
that have been arrived at when the non-Gaussian modifications are 
taken into account.
The reason behind the specific choice of the values for the parameter~$\gamma$ 
will become clear when we discuss the non-Gaussian modifications to spectra in
a subsequent subsection.


\subsection{The associated scalar bispectrum and the non-Gaussianity 
parameter}

We shall now proceed to calculate the corresponding scalar bispectra to 
eventually take into account the non-Gaussian modifications to the power 
spectra.
In scenarios involving slow roll inflation, the scalar bispectrum, say, 
$G(\vka,\vkb,\vkc)$, is known to consist of seven contributions, which 
arise from the cubic order action governing the scalar
perturbations~\cite{Maldacena:2002vr,Seery:2005wm,Chen:2010xka}. 
Of these seven contributions, six arise due to the bulk terms in the third
order action, while the seventh arises due to a field redefinition carried
out to absorb the boundary terms~\cite{Arroja:2011yj,Martin:2011sn}.
Amongst these contributions, in the situation of interest, it is known that 
the first, second, third and the seventh terms, say, $G_1(\vka,\vkb,\vkc)$, 
$G_2(\vka,\vkb,\vkc)$, $G_3(\vka,\vkb,\vkc)$ and $G_7(\vka,\vkb,\vkc)$, 
dominate the contributions due to the remaining terms.
Note that the three vectors $\vka$, $\vkb$ and $\vkc$ form the edges of a
triangle.
As we shall discuss in the following subsection, it is the bispectrum evaluated
in the so-called squeezed limit of the triangular configuration, i.e. when
$k_1\to 0$ and $k_2\simeq k_3 \simeq k$, that is expected to contribute to
the non-Gaussian modifications to the power spectrum (see, for instance,
refs.~\cite{Motohashi:2017kbs,Cai:2018dig,Unal:2018yaa}).

The scalar bispectrum in slow roll inflation with squeezed initial states can
be calculated easily using the de Sitter modes~\eqref{eq:ds-m} describing the
scalar perturbations (see, for example, refs.~\cite{Meerburg:2009ys,
Agullo:2010ws,Kundu:2011sg,Brandenberger:2012aj,Kundu:2013gha,Ganc:2011dy}).
Since the resulting expressions are somewhat lengthy, we relegate them to an 
appendix.
We have listed the complete expressions for dominant 
contributions~$G_1(\vka,\vkb,\vkc)$, $G_2(\vka,\vkb,\vkc)$, $G_3(\vka,\vkb,\vkc)$ 
and $G_7(\vka,\vkb,\vkc)$ in appendix~\ref{app:G-ab}.
It is useful to note that, in the squeezed limit, the dominant contributions to 
the scalar bispectrum at the wave number $\kf$, corresponding to the location of
the peak in the power spectrum~$\ps(k)$, can be obtained to be 
\begin{subequations}\label{eq:bs-sl-kf}
\begin{eqnarray}
\lim_{k_1\ll\kf} k_1^3\,\kf^3\,\l[G_1({\bm k}_1,{\bm k}_\mathrm{f},
-{\bm k}_\mathrm{f})
+G_3({\bm k}_1,{\bm k}_\mathrm{f},-{\bm k}_\mathrm{f})\r] 
&=& k_1^3\,\kf^3\,\l[G_1(\kf)+G_3(\kf)\r]\nn\\ 
&\simeq& \f{\HI^4}{16\,\Mpl^4\,\epsilon_1}\,\f{\kf}{k_1}\,
\f{\gamma}{\sqrt{2\,\pi\,\Delta^2_k}}\,
\l(2 + \frac{\gamma}{\sqrt{2\,\pi\,\Delta^2_k}}\r),\nn\\
\label{eq:G123_ab_sq}\\
\lim_{k_1\ll\kf} k_1^3\,\kf^3\,G_2({\bm k}_1,{\bm k}_\mathrm{f},-{\bm k}_\mathrm{f})
&=& k_1^3\,\kf^3\,G_2(\kf)\nn\\ 
&\simeq& \f{\HI^4}{16\,\Mpl^4\,\epsilon_1}\,\f{\kf}{k_1}\,
\f{\gamma}{\sqrt{2\,\pi\,\Delta^2_k}}\,
\l( 2 + \frac{\gamma}{\sqrt{2\,\pi\,\Delta^2_k}}\r),\nn\\
\label{eq:G2_ab_sq}\\
\lim_{k_1\ll \kf} k_1^3\,\kf^3\,
G_7({\bm k}_1,{\bm k}_\mathrm{f},-{\bm k}_\mathrm{f}) 
&=& k_1^3\,\kf^3\,G_7(\kf)\nn\\ 
&\simeq& \f{\HI^4\,\epsilon_2}{16\,\Mpl^4\,\epsilon_1^2}\,
\l( 1 + \frac{\gamma}{\sqrt{2\,\pi\,\Delta^2_k}}\r).\label{eq:G7_ab_sq}
\end{eqnarray}
\end{subequations}
In the above expressions, as is usually done in the context of slow roll inflation,
we have combined the contributions~$G_1(\kf)$ and~$G_3(\kf)$, as they have a similar
dependence on the wave numbers (see, for instance, ref.~\cite{Martin:2011sn}). 
We should clarify that the above expressions are the dominant contributions 
for the values of $\gamma$ we have worked with.
The striking property of the contributions $G_1(\kf)+G_3(\kf)$ and $G_2(\kf)$ is 
their dependence on the squeezed mode as~$1/k_1$. 
This property of the bispectrum in case of squeezed initial states is well
known~\cite{Agullo:2010ws,Ganc:2011dy,Brandenberger:2012aj}.
On the other hand, note that, $G_7(\kf)$ is independent of~$k_1$ in the limit
$k_1\ll \kf$. 
Therefore, at the leading order, the bispectrum around~$\kf$ is inversely 
proportional to the squeezed mode~$k_1$. 

Consider an observational survey extending over a certain range of scales such 
as, say, the measurements of the anisotropies in the CMB, which spans a few 
decades in wave numbers.
In such a case, we can calculate the squeezed limit of the bispectrum assuming~$k_1$
to be the smallest wave number within the range. 
In practice, this implies that $1\lesssim k/k_1 \lesssim 10^{4}$ over the CMB scales.
Therefore, for squeezed initial states, the bispectrum in the squeezed limit 
will be proportionately large and, hence, the associated non-Gaussianity parameter 
can be expected to be of a similar order. 
Note that, in this work, we are interested in examining phenomena leading to 
formation of PBHs and generation of secondary GWs which occur at much smaller 
scales.
For such observations spanning several decades in wave numbers, it seems 
reasonable again to choose $k_1$ to be the smallest observable wave number. 
Therefore, in our calculations, we shall set the value of squeezed mode to be 
$k_1 \simeq 10^{-4}\,\mpcinv$, which roughly corresponds to the Hubble scale 
today.
Such a choice can clearly lead to a considerable enhancement in the amplitude
of the scalar bispectrum and the corresponding non-Gaussianity parameter at the
small scales of interest.
Moreover, we should mention that, because of this boost in the amplitude, the 
consistency condition relating the scalar bispectrum to the power spectrum in 
the squeezed limit can be expected to be violated over these scales.

\begin{figure}[!t]
\centering
\includegraphics[width=15.00cm]{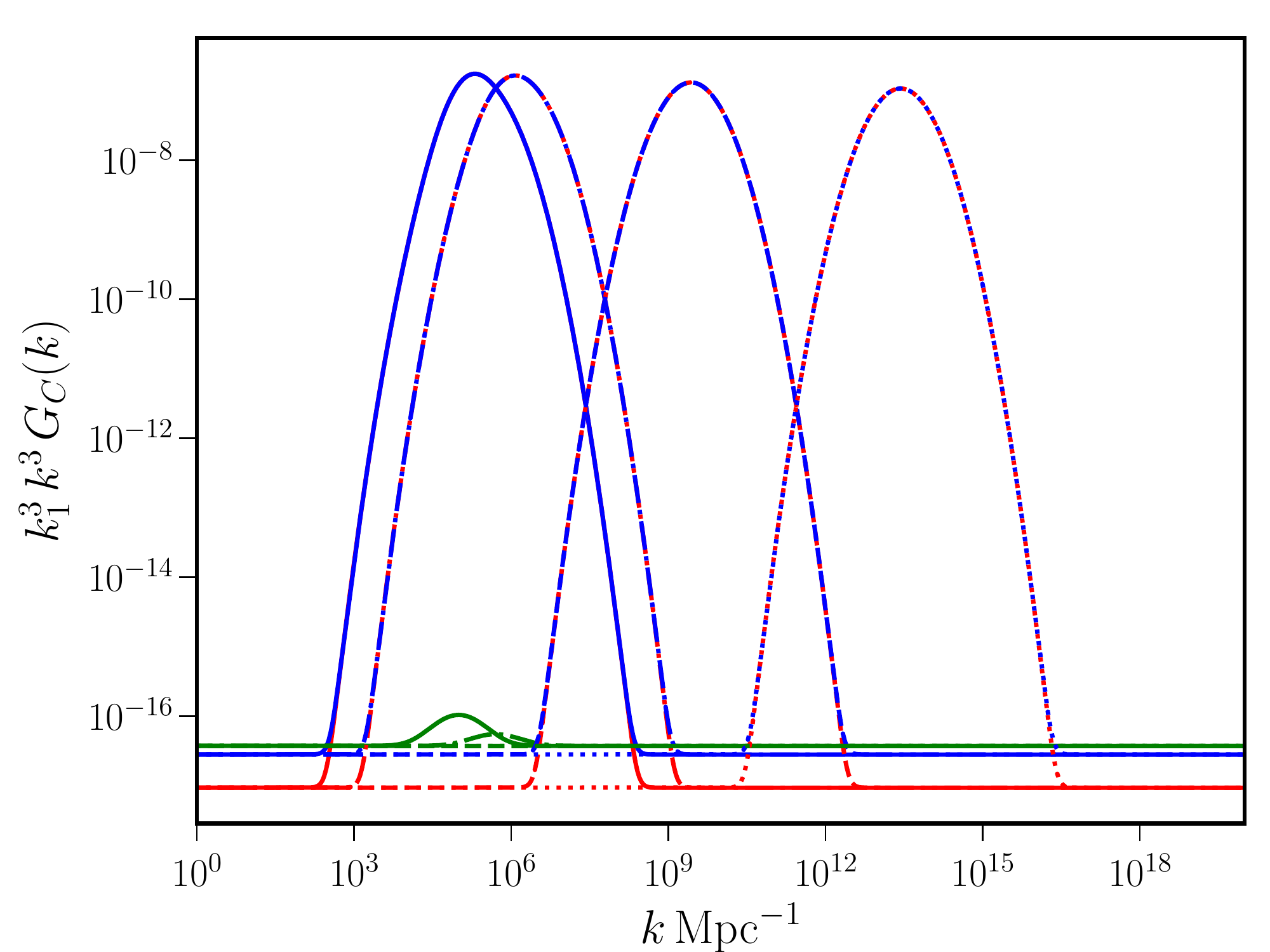}
\caption{The dominant contributions to the dimensionless scalar bispectra 
in the squeezed limit, viz. $k_1^3\,k^3$ times $G_1(k) + G_3(k)$, $G_2(k)$ 
and $G_7(k)$, have been plotted (in red, blue and green, respectively) for 
non-vacuum initial states which lead to scalar power spectra with lognormal 
peaks. 
We have plotted the contributions to the dimensionless bispectra for the four 
sets of values for the parameters $\gamma$ and $\kf$ (as solid, dashed-dotted, 
dashed and dotted curves) we had considered in the previous figure. 
It is clear that the bulk terms $G_C(k)$ with $C=\{1,2,3\}$ dominate the 
contributions to the bispectrum.
In contrast, as expected, the boundary term $G_7(k)$ has a much smaller 
amplitude and mimics the shape of the power spectrum.}\label{fig:k3k3G}
\end{figure}

In figure~\ref{fig:k3k3G}, we have plotted the behavior of the bispectrum in 
the squeezed limit for the four set of values for the parameters of~$\gamma$ 
and~$\kf$ we considered earlier.
Notice that the amplitudes of the bispectra are significantly enhanced around 
the locations of the peaks in the power spectra. 
The amplitudes retain their slow roll values away from the peaks. 
The amplification of several orders of magnitude around $\kf$ arises evidently 
due to the dependence of the bispectrum on the squeezed mode as $1/k_1$, as we
discussed above. 
We should stress that this amplification occurs even for a relatively small value
of the parameter~$\gamma$, which quantifies the deviations from the Bunch-Davies
vacuum.
We find that, for a larger $k$, we require a smaller value of $\gamma$ to achieve
the same level of enhancement of the bispectrum.
In other words, the bispectrum becomes increasingly sensitive to deviations from 
the standard vacuum state at smaller scales.

\begin{figure}[!t]
\centering
\includegraphics[width=15.00cm]{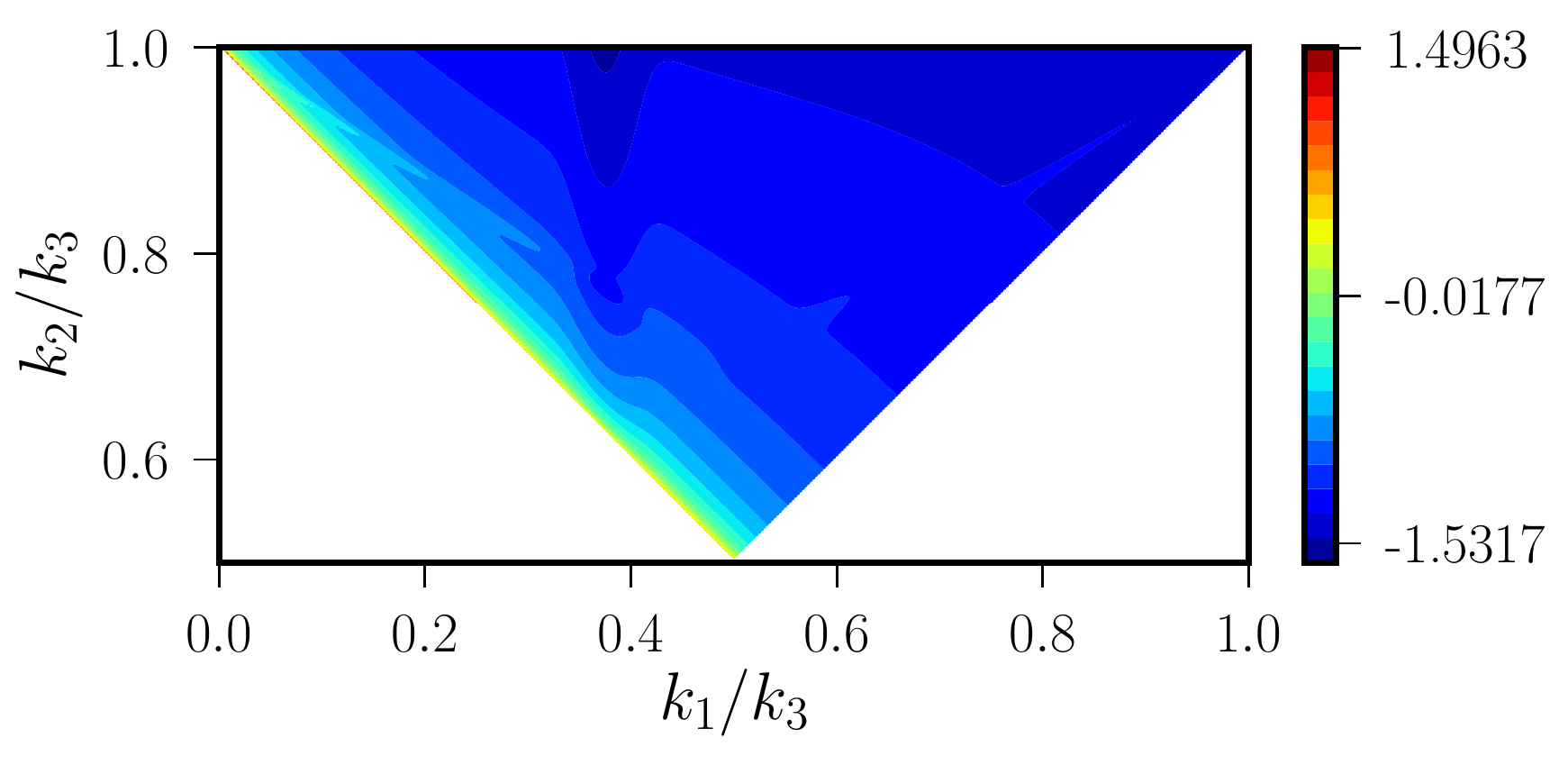}
\caption{The non-Gaussianity parameter $\mathrm{log}\,\vert \fnl \vert$ has 
been plotted as a density plot in the $k_1/k_3$--$k_2/k_3$ plane, for the 
first of the four sets of parameters we had introduced in figure~\ref{fig:ps}. 
We have set $k_3=\kf$ and varied $k_1/k_3$ over the range $[5\times 10^{-4},1]$ 
in arriving at this figure. 
Note that the $\fnl$ parameter has a largely `local' shape, with its maximum 
amplitude (in red) occurring in the so-called flattened limit corresponding to 
the left edge of the triangle.}\label{fig:s-of-fnl}
\end{figure}

The non-Gaussianity parameter associated with the scalar 
bispectrum~$G(\vka,\vkb,\vkc)$ is defined as~\cite{Martin:2011sn,Hazra:2012yn}
\begin{eqnarray}
\fnl(\vka,\vkb,\vkc)
& =&-\f{10}{3}\,\f{1}{\l(2\,\pi\r)^4}\;k_1^3\, k_2^3\, k_3^3\;
G(\vka,\vkb,\vkc)\nn\\
& & \times\, \biggl[k_1^3\,\ps(k_2)\,\ps(k_3) 
+ {\mathrm{two~permutations}}\biggr]^{-1}.\label{eq:fnl}
\end{eqnarray}
The dimensionless parameter $\fnl(\vka,\vkb,\vkc)$ can be calculated using the 
expressions~\eqref{eq:gk}, \eqref{eq:gk-ln} and~\eqref{eq:G-ce} for the power
spectrum, the function $g(k)$ and the bispectrum. 
In order to understand the complete shape of the scalar bispectrum, 
in figure~\ref{fig:s-of-fnl}, we have illustrated the non-Gaussianity 
parameter as a density plot in the $k_1/k_3$--$k_2/k_3$~plane for
the first of the four sets of parameters for~$\gamma$ and $\kf$ we
had introduced earlier (see the caption of figure~\ref{fig:ps}).
The figure clearly illustrates the fact that the non-Gaussianity parameter 
has a largely `local' shape.
As is well known, its amplitude is the largest in the flattened limit, 
i.e. along the line $k_2/k_3=1-k_1/k_3$ which describes the left edge
of the triangle in the figure~\ref{fig:s-of-fnl}.
This shape evidently depends on the choice of~$k_3$, which in this 
illustration has been set to be the location of the peak~$\kf$.

\begin{figure}[!t]
\centering
\includegraphics[width=15.00cm]{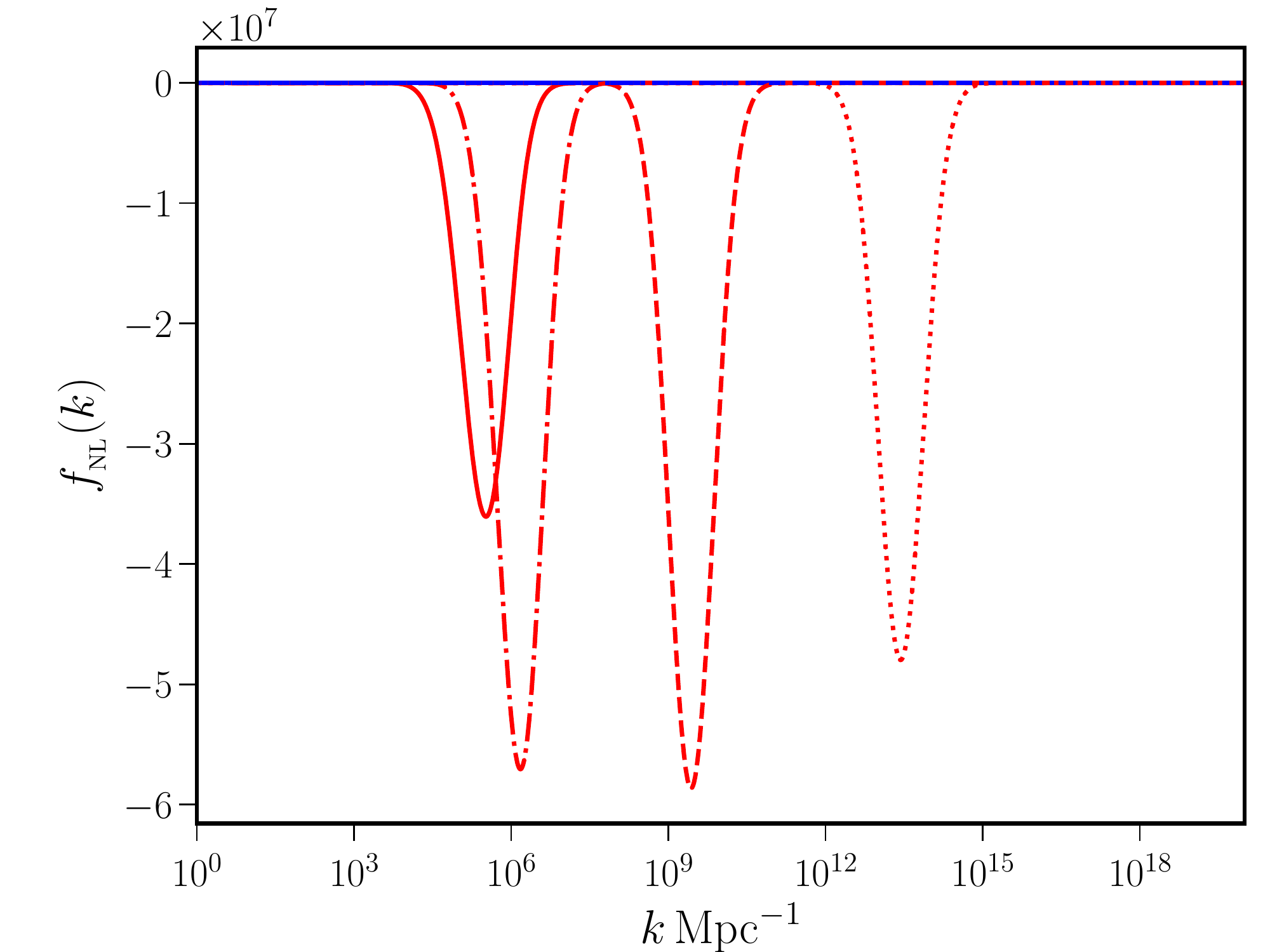}
\caption{The non-Gaussianity parameter $\fnl(k)$ in the squeezed limit has 
been plotted (in red) for the four set of parameters (as solid, dashed-dotted, 
dashed and dotted curves) leading to lognormal spectra we had considered in 
the first two figures.
We have also plotted the quantity~$\fnl^{^{_{\mathrm{CR}}}}(k)$ (in blue) for
all the cases to illustrate the fact that the consistency condition is strongly 
violated around the region of the peaks in the power spectra.}\label{fig:fnl-sl}
\end{figure}

Let us now turn to consider the behavior of the parameter $\fnl$ in the 
squeezed limit.
In such a limit, on utilizing the results~\eqref{eq:bs-sl-kf}, we obtain the
value of $\fnl$ at the location of the peak in the power spectrum~$\ps(k)$ 
to be
\begin{equation}
\lim_{k_1\ll\kf}
\fnl^{^{_{\mathrm{SL}}}}({\bm k}_1,{\bm k}_\mathrm{f},-{\bm k}_\mathrm{f})
=\fnl^{^{_{\mathrm{SL}}}}(\kf)
\simeq -\f{5\,\epsilon_1}{6}\,\f{\kf}{k_1}
\f{\gamma}{\sqrt{2\,\pi\,\Delta_k^2}}\,
\l(\f{2 + \f{\gamma}{\sqrt{2\,\pi\,\Delta_k^2}}}{1 + \f{\gamma}{\sqrt{2\,\pi\,\Delta_k^2}}}\r).\label{eq:fnl_ab_sq}
\end{equation}
In figure~\ref{fig:fnl-sl}, we have plotted the behavior of $\fnl(k)$ in the 
squeezed limit for the four sets of parameters we have mentioned earlier.
We find that, for these choices of the parameters, the value of $\fnl$ is of
order $10^7$ around~$\kf$, while it has the slow roll value of $10^{-2}$ away 
from~$\kf$. 
Also, we find that the consistency condition~---~viz. that 
$\fnl^{^{_{\mathrm{CR}}}}(k) = 5\, \l[\ns(k)-1\r]/12$, where 
$\ns(k)=1+\d\, \mathrm{ln}\,\ps(k)/\d\, \mathrm{ln}\, k$ is the 
scalar spectral index~---~is strongly violated around the 
lognormal peak as expected, while it is satisfied sufficiently
far away from the peak.
It has been argued that any calculation of $\fnl$ has to account 
for the so-called local observer effect (in this context, see, 
for instance, refs.~\cite{Tada:2016pmk,Suyama:2020akr}).
This essentially means that, to arrive at the observable value of the 
non-Gaussianity parameter in the squeezed limit, we need to subtract 
the part of~$\fnl$ satisfying the consistency relation from its total 
value.
In the scenario of interest, around the peaks in the power spectra,
the quantity $\fnl^{^{_{\mathrm{CR}}}}(k)$ is negligible compared 
to the magnitude of the~$\fnl$ obtained from the squeezed initial 
states.
The main conclusions we can draw from the above considerations are twofold.
Firstly, for perturbations evolved from non-vacuum initial states, the
non-Gaussianity parameter~$\fnl$ is inversely proportional to the value 
of squeezed mode.
Hence, it has a rather large amplitude over small scales for 
the values of the parameter $\gamma$ we have considered.
Secondly, the amplitude of $\fnl$ is highly sensitive to even minor 
deviations from standard vacuum state.
As we shall discuss in the following subsection, the large value for the 
non-Gaussianity parameter in the squeezed limit leads to substantial
modifications to the original power spectrum.
This should be contrasted with scenarios involving, say, ultra slow
roll inflation, wherein the consistency condition governing the 
scalar bispectrum is satisfied in the squeezed limit and hence the 
non-Gaussian corrections to the power spectrum prove to be either 
negligible or identically zero~\cite{Ragavendra:2020sop,Bravo:2020hde}.


\subsection{Non-Gaussian modifications to the scalar power spectrum}

Having arrived at the bispectrum and the corresponding non-Gaussianity
parameter, let us now proceed to calculate the non-Gaussian modification
to the scalar power spectrum~\cite{Garcia-Bellido:2017aan,Cai:2018dig,
Unal:2018yaa,Yuan:2020iwf,Ragavendra:2020sop}.
Recall that the non-Gaussianity parameter~$\fnl$ is usually introduced 
through the following relation (see ref.~\cite{Komatsu:2001rj}; also see,
for example, refs.~\cite{Martin:2011sn,Hazra:2012yn}):
\begin{equation}
\cR(\eta, \vx)=\cR^{^{_{\mathrm{G}}}}(\eta, \vx)
-\f{3}{5}\,\fnl\, \l[\cR^{^{_{\mathrm{G}}}}(\eta, \vx)\r]^2,
\label{eq:i-fnl}
\end{equation}
where $\cR$ is the scalar perturbation and $\cR^{^{_{\mathrm{G}}}}$ 
denotes the Gaussian contribution. 
In Fourier space, this relation can be written as (see, for instance, 
refs.~\cite{Martin:2011sn,Unal:2018yaa})
\begin{equation}
\cR_\vk
=\cR^{^{_{\mathrm{G}}}}_\vk
-\f{3}{5}\,\fnl\, \int \f{\d^{3}{\vp}}{(2\,\pi)^{3/2}}\, 
\cR^{^{_{\mathrm{G}}}}_{\vp}\; 
\cR^{^{_{\mathrm{G}}}}_{\vk-\vp}.
\end{equation}
If one uses this expression for~$\cR_\vk$ and evaluates the corresponding
two-point correlation function in Fourier space, one obtains 
that~\cite{Cai:2018dig,Unal:2018yaa,Ragavendra:2020sop} 
\begin{equation}
\langle \hat{\cR}_\vk\, \hat{\cR}_{\vk'}\rangle
=\f{2\,\pi^2}{k^3}\,\delta^{(3)}(\vk +\vk')\,
\l[\ps(k)+\l(\f{3}{5}\r)^2\,\f{k^3}{2\,\pi}\,\fnl^2\,
\int {\rm d}^{3}{\vp}\,
\f{\ps(p)}{p^3}\,\f{\ps\l(\vert \vk-\vp\vert\r)}{\vert\vk-\vp\vert^3}\r],
\label{eq:psc}
\end{equation}
where $\ps(k)$ is the original scalar power spectrum defined in the Gaussian 
limit, while the second term represents the leading non-Gaussian modifications.
It can be easily shown that the non-Gaussian modification to the scalar power 
spectrum, say, $\pc(k)$, can be expressed as 
\begin{eqnarray}
\pc(k)&=&\l(\f{12}{5}\r)^2\,\fnl^2\,
\int_{0}^{\infty}\d s\int_{0}^{1}\f{\d d}{(s^2-d^2)^2}\,
\ps[k\,(s+d)/2]\, \ps[k\,(s-d)/2].\label{eq:pc-k}
\end{eqnarray}

We should clarify a few points at this stage of our discussion.
We should mention that the quantity $\fnl$ has been assumed to be local 
in arriving at the above expression for the correction to the power 
spectrum~$\pc(k)$. 
Therefore, we shall work with the value $\fnl$ in the 
squeezed limit when calculating the non-Gaussian modifications to the 
power spectrum.
(Recall that, around~$\kf$, the scalar bispectrum had a largely `local' shape, 
as illustrated in figure~\ref{fig:s-of-fnl}.)
Moreover, the parameter $\fnl$ in the squeezed limit in our scenario is 
highly scale dependent in the sense that it is large around~$\kf$ 
(for the values of the parameter $\gamma$ we have worked with), but is
completely negligible away from it.
Hence, when calculating the modifications to the spectrum, in 
eq.~\eqref{eq:pc-k}, we have assumed $\fnl$ to be a function of $k$.
In figure~\ref{fig:ps}, we have plotted the modified spectra, viz. $\ps(k)
+\pc(k)$, as well as the spectra $\ps(k)$ we had originally constructed.
Note that the non-Gaussian modifications $\pc(k)$ dominate at small scales
around the peaks in the original power spectra. 
In fact, it is due to the dependence of the non-Gaussianity parameter~$\fnl$ 
on the squeezed mode as $1/k_1$ that we have been able to achieve the required 
boost in the power spectrum [of $\mathcal{O}(10^{-2})$] at small scales.
Also, we should point out that, given a $\gamma$, the amplification due to 
the non-Gaussian modifications are larger at a higher~$\kf$.
It is due to this reason that, for a larger $\kf$, we have worked with a 
smaller value of~$\gamma$.
We have chosen these parameters so that, when the non-Gaussian modifications
are taken into account, the modified power spectra have comparable amplitudes
at their maxima despite the varying amplitudes of the peaks in their original 
spectra. 
We should clarify that the large, cubic order, non-Gaussian 
corrections do not lead to a breakdown of the perturbation theory since the 
scalar power spectra are of $\mathcal{O}(10^{-2})$ even when the modifications 
due to the scalar bispectra have been taken into account (cf. figure~\ref{fig:ps}).

It is worthwhile to highlight another related point at this stage of our 
discussion.
We find that the widths of the modified power spectra are larger than the 
widths of the original power spectra which were dictated by the parameter
$\Delta_k$ that we have set to unity.
This is because of the nature of the integrand involved that describes the 
non-Gaussian correction given in eq.~\eqref{eq:pc-k}. 
The appearance of the integration variables~$s$ and~$d$ in the arguments of 
the original power spectrum as well as the limits of the integrals involved 
contribute to the widening of the peak and a slight shift of power towards
larger wave numbers in the final modified spectra.


\section{Formation of PBHs and generation of secondary GWs}\label{sec:o}

In this section, we shall calculate the observable quantities~$\fpbh(M)$
and~$\ogw(f)$ using the scalar power spectra with the non-Gaussian 
corrections taken into account.

Given a primordial scalar power spectrum~$\ps(k)$, there exists a standard 
procedure to arrive at the corresponding~$\fpbh(M)$ characterizing the 
fraction of PBHs constituting dark matter today.
Let us quickly recall the essential points in this regard.
We shall focus on scales that reenter the Hubble radius during the radiation
dominated epoch.
In such a case, the observable $\fpbh$ can be expressed in terms of the 
mass $M$ of the PBHs as follows (in this context, see the
reviews~\cite{Carr:2016drx,Carr:2018rid,Sasaki:2018dmp,Carr:2020xqk}):
\begin{equation}
\fpbh(M) 
=\l(\f{\gamma_\ast}{0.2}\r)^{3/2}\,
\l(\f{\beta(M)}{1.46\times 10^{-8}}\r)\, 
\l(\f{g_{\ast,k}}{g_{\ast,\mathrm{eq}}}\r)^{-1/4}\,
\l(\f{M}{M_\odot}\r)^{-1/2},\label{eq:fpbh-f}
\end{equation}
where $\beta(M)$ denotes the fraction of the energy density of PBHs to 
the total energy density of the universe at the time of their formation.
The quantities $g_{\ast,k}$ and $g_{\ast,{\mathrm{eq}}}$ are the number 
of effective relativistic degrees of freedom at the time of formation of 
the PBHs and at matter-radiation equality, respectively, while $\gamma_\ast$ 
denotes the efficiency of the process leading to the formation of 
black holes. 
We shall set $g_{\ast,k} = 106.75\,$, $g_{\ast,{\mathrm{eq}}}=3.36$ and 
$\gamma_\ast=0.2$, as is often done in this context.
If we now assume that perturbations beyond a threshold density contrast,
say, $\delta_\mathrm{c}$, are responsible for the formation of PBHs, 
then the function $\beta(M)$ is given by 
\begin{eqnarray}
\beta(M) 
\simeq 
\f{1}{2}\,\l[1-\mathrm{erf}\l(\f{\delta_\mathrm{c}}{\sqrt{2\,\sigma^2(R)}}\r)\r],
\label{eq:b-pbh}
\end{eqnarray}
where $\mathrm{erf}(z)$ is the error function.
The variance $\sigma^2(R)$ is related to the primordial scalar power 
spectrum~$\ps(k)$ through the integral
\begin{equation}
\sigma^2(R) 
= \f{16\,R^4}{81}\int_{0}^{\infty} \d\,{\mathrm{ln}}k\;k^4\,
\ps(k)\, W^2(k,R),\label{eq:sigma2}
\end{equation}
where $W(k,R)$ is a window function with a smoothening radius~$R$, 
which we shall assume to be a Gaussian of the form $W(k\,R) = 
\mathrm{e}^{-(k^2\,R^2)/2}$.
Note that the length scale $R$ is related to the mass~$M$ of PBHs 
through the expression
\begin{equation}
R=4.72\times10^{-7}\,\l(\f{\gamma}{0.2}\r)^{-1/2}\,
\l(\f{g_{\ast,k}}{g_{\ast,\mathrm{eq}}}\r)^{1/12}\,
\l(\f{M}{M_\odot}\r)^{1/2}\,\mathrm{Mpc}\,.\label{eq:R-Ms}
\end{equation}
Therefore, given a power spectrum $\ps(k)$ we can first compute 
the variance~$\sigma^2(R)$. We should clarify that we shall make use of the 
scalar power spectrum with the non-Gaussian modifications taken into account, 
i.e. we shall consider $\ps(k)+\pc(k)$.
We can then make use of the above relation between $R$ and $M$ 
and the expression for~$\beta(M)$ to finally arrive at $\fpbh(M)$ 
utilizing eq.~\eqref{eq:fpbh-f}. 
It is well known that the threshold of the density 
contrast~$\delta_{\mathrm{c}}$ is a crucial parameter 
since $\fpbh$ is exponentially sensitive to it. 
The value of~$\delta_{\mathrm{c}}$ is expected to lie in the range $0.3$--$0.65$ 
(see refs.~\cite{Musco:2018rwt,Escriva:2019phb,Kehagias:2019eil,Escriva:2020tak}, 
see however the recent discussion~\cite{Musco:2020jjb}).
For the purposes of illustration, we shall work with $\delta_{\mathrm{c}}=1/3$ 
and $0.5$.
We should clarify that the exact value of this parameter does not affect the
primary conclusions we draw about the mechanism of generating PBHs from 
squeezed initial states.

As we mentioned, the amplification of scalar power at small scales invariably
produces secondary GWs of significant strength as they are sourced by the second order scalar perturbations~\cite{Ananda:2006af,Baumann:2007zm,Saito:2008jc,
Saito:2009jt}.
With the scalar power spectra obtained from squeezed initial states, we shall 
also proceed to calculate the dimensionless energy density $\ogw$ of the
secondary GWs today as a function of the frequency, say,~$f$. 
The calculations involved are well understood~\cite{Kohri:2018awv,
Espinosa:2018eve,Domenech:2019quo,Pi:2020otn,Domenech:2020kqm}. 
We should mention here that, as we had done in the calculation of $\fpbh(M)$, 
we shall take into account the non-Gaussian modifications to the power spectrum 
to arrive at~$\ogw(f)$~\cite{Cai:2018dig,Unal:2018yaa}.
Recall that we are focusing on scales that reenter the Hubble radius during
the radiation dominated epoch. 
In such a case, the second order tensor perturbations induced by the scalar
perturbations oscillate in the sub-Hubble regime.
Upon averaging over small time scales corresponding to these oscillations, the 
power spectrum of the secondary tensor perturbations, say, $\overline{\ph(k,\eta)}$, 
can be expressed in terms of the scalar power spectrum as 
follows (see, for instance, refs.~\cite{Bartolo:2016ami,Bartolo:2018evs,
Bartolo:2018rku,Espinosa:2018eve}):
\begin{eqnarray}
\overline{\ph(k,\eta)}
&=& \f{2}{81\,k^2\,\eta^2}
\int_{0}^{\infty}\d v\,\int_{\vert 1-v\vert}^{1+v}\d u\,
\l[\f{4\,v^2-(1+v^2-u^2)^2}{4\,u\,v}\r]^2\,\ps(k\,v)\,\ps(k\,u)\nn\\
& &\times\,\l[\cI_c^2(u,v)+\cI_s^2(u,v)\r],\label{eq:phf}
\end{eqnarray}
where the functions $\cI_c(u,v)$ and $\cI_s(u,v)$ 
are given by~\cite{Kohri:2018awv,Espinosa:2018eve}
\begin{subequations}\label{eq:cI}
\begin{eqnarray}
\cI_c(v,u) &=& -\f{27\,\pi}{4\,v^3\,u^3}\,
\Theta\l(v+u-\sqrt{3}\r)\, (v^2+u^2-3)^2,\\
\cI_s(v,u) &=& -\f{27}{4\,v^3\,u^3}\, (v^2+u^2-3)\,
\l[4\,v\,u+ (v^2+u^2-3)\;{\rm log}\,
\biggl\vert\f{3-(v-u)^2}{3-(v+u)^2}\biggr\vert\r]
\end{eqnarray}
\end{subequations}
with $\Theta(z)$ denoting the step function.
The dimensionless energy density associated with the secondary
GWs~$\ogw(k,\eta)$, evaluated at late enough times when the modes 
are inside the Hubble radius during the radiation dominated epoch, 
is given by
\begin{equation}
\ogw(k,\eta)
=\f{1}{24}\,\l(\f{k}{{a\,H}}\r)^2\, \overline{\ph(k,\eta)}.\label{eq:ogw-rd}
\end{equation}
The observable quantity of interest, viz. the energy density of secondary 
GWs evaluated today~$\ogw(f)$ (with~$f$ being the frequency associated with the 
wave number~$k$), can be written in terms of the quantity~$\ogw(k,\eta)$ 
above as 
\begin{equation}
h^2\,\ogw(k)
\simeq 1.38\times10^{-5}\, 
\l(\f{g_{\ast,k}}{106.75}\r)^{-1/3}\,
\l(\f{\Omega_{\mathrm{r}}\,h^2}{4.16\times10^{-5}}\r)\,\ogw(k,\eta),\label{eq:ogw0}
\end{equation}
where $\Omega_{\mathrm{r}}$ denotes the present day dimensionless energy 
density of relativistic matter and $h$ is the usual parameter introduced 
to describe the Hubble parameter today as $H_0=100\,h\,\mathrm{km}\,
\mathrm{s}^{-1}\,\mpcinv$.

\begin{figure}[!t]
\centering
\includegraphics[width=7.50cm]{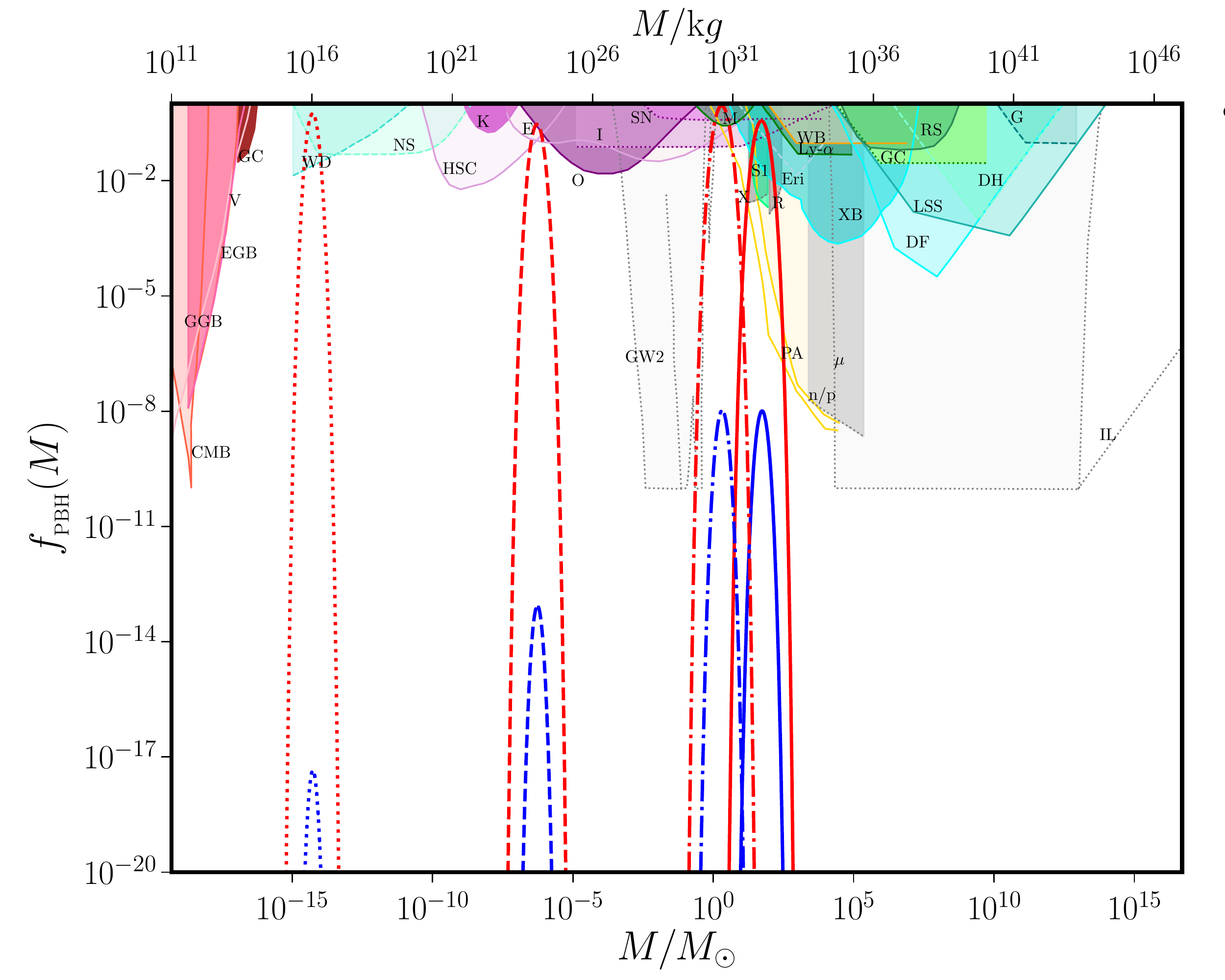}
\includegraphics[width=7.50cm]{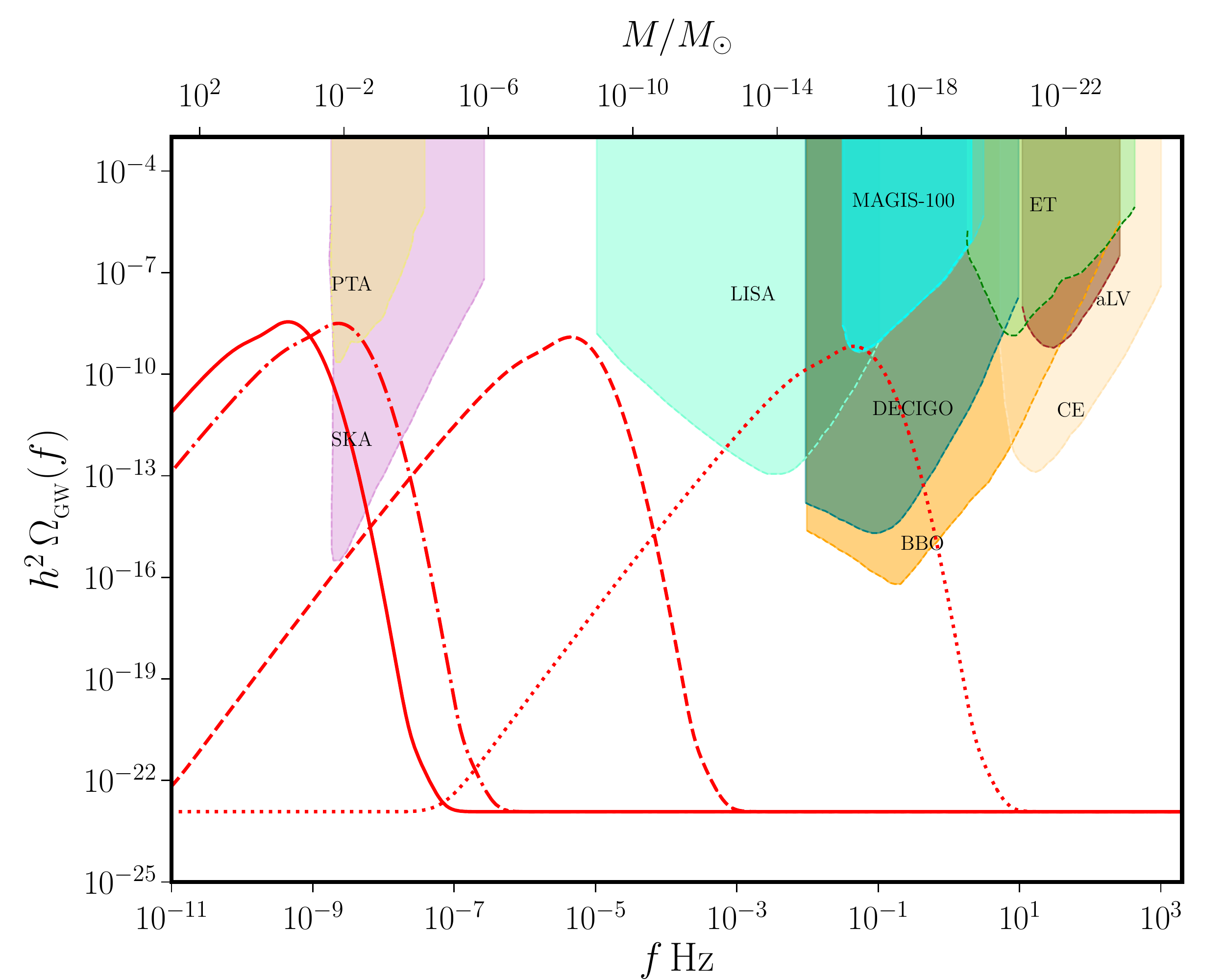}
\vskip -10pt
\caption{The quantity~$\fpbh(M)$ (on the left, for $\delta_\mathrm{c}=1/3$ 
and $0.5$ in red and blue, respectively) and the dimensionless energy density 
of GWs~$\ogw(f)$ (on the right) have been plotted for the cases of the four 
lognormal spectra with the non-Gaussian modifications to the power spectrum 
taken into account that were illustrated in figure~\ref{fig:ps}.
The various constraints on $\fpbh(M)$ from different observations have 
also been indicated (on top of the figure on the left) over the 
corresponding mass ranges.
We have also included the sensitivity curves of the various ongoing and 
upcoming observational missions of GWs (as shaded regions in the top part
of the figure on the right).
The intersections of the curves with the shaded regions translate to constraints 
on the parameter~$\gamma$ which determines the extent of deviation of the 
initial state from the Bunch-Davies vacuum.}\label{fig:fpbh-sgw}
\end{figure}

In figure~\ref{fig:fpbh-sgw}, we have plotted the quantities $\fpbh(M)$ 
and $\ogw(f)$ for the four power spectra we have obtained from squeezed 
initial states with the non-Gaussian modifications taken into account 
[cf. eqs.~\eqref{eq:psc} and~\eqref{eq:pc-k}].
We have also included the constraints on $\fpbh(M)$ that are presently 
available from different datasets in the various mass
ranges (see refs.~\cite{Carr:2009jm,Carr:2016drx}; 
for recent discussions of the constraints over 
specific mass ranges, see refs.~\cite{Dasgupta:2019cae,Unal:2020mts}). 
Moreover, we have illustrated the sensitivity curves of the various GW 
observatories and missions (in this context, see ref.~\cite{Moore:2014lga}). 
As expected, the enhancements in the scalar power on small scales lead 
to proportional amplifications in $\fpbh(M)$ and $\ogw(f)$ over the 
corresponding masses and frequencies.
Also, due to the nature of the integrals that determine~$\overline{\ph(k,\eta)}$
[cf. eq.~\eqref{eq:phf}], the peaks of $\ogw(f)$ are considerably wider when 
compared to the peaks of the scalar power spectra.
As can be seen from the figure, the predicted $\fpbh(M)$ and $\ogw(f)$ curves
already intersect the various constraints and sensitivity curves. 
These constraints immediately translate to bounds on the parameter~$\gamma$
which determines the strength of the feature in the scalar power spectra. 
Recall that, the Bogoliubov coefficient~$\beta(k)$ is proportional to $\gamma$ 
[cf. eq.~\eqref{eq:al-be}]. 
So, in our scenario of PBHs and secondary GWs produced from excited initial states,
evidently, the limits on $\fpbh$ and $\ogw$ directly constrain the non-vacuum nature 
of the states from which the perturbations evolve. 


\section{Challenges associated with squeezed initial states}\label{sec:c}

In the last two sections, we have illustrated that a specific choice for 
the Bogoliubov coefficient~$\beta(k)$ can lead to the desired lognormal
peak in the scalar power spectrum [cf.~eqs.~\eqref{eq:gk}, \eqref{eq:gk-ln}
and \eqref{eq:al-be}].
We have also shown that, since the cubic order non-Gaussian corrections prove 
to be significant in the squeezed limit in the non-vacuum initial states, it 
is possible to choose a relatively small value for $\beta(k)$ to arrive at
large peaks in the effective scalar power spectrum. 
We have also examined the possible imprints of such power spectra on the
extent of PBHs produced and the secondary GWs generated on small scales.
In this section, we shall discuss some of the challenges associated with 
squeezed initial states.


\subsection{Possible mechanisms to generate squeezed states}

The first task before us is to justify the choice of the squeezed initial 
states of our interest.
In other words, we need to examine whether there exist mechanisms that 
can generate the specific form of~$\beta(k)$ that we have considered.
Note that, we have assumed that the curvature perturbation is in the
non-vacuum initial state at some early time, say, $\ei$, when the 
smallest wave number of our interest, viz. $k_1 \simeq 10^{-4}\,
\mpcinv$, is adequately inside the Hubble radius.
In this subsection, we shall discuss mechanisms that can possibly excite
the curvature perturbations to such an initial state and the challenges 
associated with them.

The first possibility would be to consider effects due to high energy physics. 
For instance, since the large scale modes emerge from sub-Planckian length 
scales during the initial stages of inflation, it has been argued that 
trans-Planckian physics may modify the dynamics of the perturbations during
the early stages (for the original discussion, see 
ref.~\cite{Brandenberger:2002hs}).
But, in the absence of a viable model of quantum gravity to take into account 
the high energy effects, the equations describing the perturbations are often 
modified by hand.
The modifications essentially introduce an energy scale into the equations of
motion governing the perturbations, beyond which the new physics operates, while 
ensuring that the standard equations are satisfied at lower energies.
One of the approaches that has been extensively examined in this context 
involves modifying the dispersion relation governing the perturbations (for
example, see the review~\cite{Brandenberger:2012aj}).
In this context, while the super-luminal dispersion relations
are known to leave the primordial spectrum largely unaffected,
the sub-luminal dispersion relations have been shown to lead to 
significant production of particles resulting in stronger features 
in the power spectrum~\cite{Brandenberger:2012aj}.
However, the produced particles result in significant backreaction (a point which
we shall discuss in the following subsection) making them unviable.
We also find that, in some of the approaches, the power spectrum is modified on 
large scales, since they emerge from the sub-Planckian length scales at high 
energies (see, for instance, ref.~\cite{Shankaranarayanan:2004iq}).
Another popular method that has been considered to take into account the high 
energy effects involves the imposition of non-trivial initial conditions on the 
standard modes as they emerge from the Planckian regime~\cite{Danielsson:2002kx}.
Such an approach is known to only result in oscillations in the power spectrum 
over a wide range of scales~\cite{Martin:2003sg,Martin:2004yi}.

Another possibility that can leave the curvature perturbation in an excited 
state during the early stages of inflation would be to consider an initial 
epoch of non-inflationary phase.
Often, one either considers a radiation dominated phase or an initial period
wherein the scalar field is rolling rapidly (in this context, see, for 
instance, refs.~\cite{Powell:2006yg,Contaldi:2003zv}; for recent discussions, 
see refs.~\cite{Hergt:2018ksk,Ragavendra:2020old}).
Again, in such cases, the power spectrum seems to be modified only on large 
scales and it often displays a sharp drop in power over these scales.
Moreover, we should add that, in such scenarios, it is possible that a certain 
range of wave numbers would have never been inside the Hubble radius.
Therefore, there can arise some ambiguity in the initial conditions that are
to be imposed on these modes.
Moreover, we should mention that, if such a pre-inflationary mechanism is to 
excite the state of the curvature perturbation at the small wave numbers~$\kf$ 
of interest, the mechanism should involve changes that occur as rapidly 
as $\kf^{-1}$~(for a recent related discussion, see, for example, 
ref.~\cite{Hashiba:2021npn}). 
Yet another possibility would be to consider two stages of slow roll inflation
with either a brief departure from slow roll or even a break from inflation 
sandwiched between them.
But, these are exactly the scenarios of ultra slow roll and punctuated inflation 
that have been considered to generate increased power on small scales 
so as to lead to enhanced formation of PBHs and higher strengths of secondary
GWs~\cite{Germani:2017bcs,Bhaumik:2019tvl,Ragavendra:2020sop,Bhaumik:2020dor}). 
Apart from single field models, as had mentioned in the introductory section, there
also exist inflationary scenarios involving two fields which can lead to a rapid rise
in power on small scales~\cite{Braglia:2020eai,Fumagalli:2020nvq,Braglia:2020taf}.
Often, in this context, there arises a sharp turn in the trajectory of the 
fields, essentially giving rise to particle production and therefore a 
non-trivial form of~$\beta(k)$ (in this context, see the discussion in
ref.~\cite{Fumagalli:2020nvq}).  
However, these models involve a certain level of fine tuning of the field 
trajectory and the form of~$\beta(k)$ will be dependent on the details of 
the model.
Importantly, we should mention that, in such cases, the features are generated 
as the modes of interest leave the Hubble radius during the epochs of deviations
from slow roll.
Actually, this is true of any inflationary scenario. 
This implies that it is difficult to generate features on small scales as we
desire by inducing or introducing transitions in or between inflationary phases 
at very early stages.

In fact, there exists one more possibility.
One can treat the curvature perturbation that we are considering as associated 
with a test field in an inflationary regime driven by another source (for
scenarios wherein the dominating background is driven by another scalar field,
see, for instance, refs.~\cite{Felder:1999wt,Wang:2013oea}; for situations 
wherein the perturbations are dominated by, say, the Higgs field, see
refs.~\cite{Lu:2019tjj,Espinosa:2018eve}).
The source that dominates the background dynamics either prior to inflation or 
in the early stages of the inflationary regime can excite the modes associated
with the curvature perturbations leaving it in a squeezed state.
Let us illustrate the points we wish to make in this regard by starting with
the aid of an example.
Consider a situation wherein the Fourier mode $\psi_k$ of a quantum field 
satisfies an equation of motion of the following form:
\begin{equation}
\psi_k''+\l(k^2+\mu^2\,k_0^2\,\eta^2\r)\,\psi_k=0,\label{eq:de-i}
\end{equation}
where~$\mu$ and~$k_0$ denote scales associated with the system.
The solution to such a differential equation can be expressed in terms of 
the parabolic cylinder functions and by comparing the asymptotic forms of 
the solutions at early and late times, one can immediately show that the 
number of particles produced in such a case is given by (in this context, 
see the discussions in the recent work~\cite{Hashiba:2021npn})
\begin{equation}
\vert\beta(k)\vert^2=\mathrm{e}^{-k^2/(\mu\,k_0)}.\label{eq:beta-se}    
\end{equation}
In fact, such a result should not come as a surprise.
One encounters an equation of motion of the above form when one considers a 
complex scalar field that is evolving in the background of a constant electric 
field in flat spacetime, leading to the well known Schwinger
effect~\cite{Schwinger:1951nm}.
Note that the above Bogoliubov coefficient (to be precise, its modulus squared) 
is a Gaussian, which is close to the form that we desire.
However, since it is not of the lognormal shape, it is peaked at $k=0$ rather 
than at a non-zero~$k$.
Moreover, it has a maximum value of unity, whereas we require an additional 
parameter (such as $\gamma$) to be able to tune the amplitude of~$\beta(k)$.

Let us now discuss mechanisms that can possibly help us achieve the 
desired~$\beta(k)$ in a FLRW universe.
A good starting point seems to be to construct situations in which the 
equation governing either the curvature perturbation or a test scalar 
field has the same form as eq.~\eqref{eq:de-i} above so that we can at 
least arrive at a Gaussian form for $\vert \beta(k)\vert^2$. 
Recall that the Mukhanov-Sasaki variable $v_k$ associated with the curvature
perturbation satisfies the equation
\begin{equation}
v_k''+\l(k^2-\f{z''}{z}\r)\,v_k=0,
\end{equation}
where $z=\sqrt{2\,\epsilon_1}\,\Mpl\,a$.
Evidently, we require $z''/z=-\mu^2\,k_0^2\,\eta^2$ if we are to achieve 
the $\vert\beta(k)\vert^2$ mentioned above [cf. eq.~\eqref{eq:beta-se}].
In such a case, the generic solution to $z$ can be immediately expressed
in terms of a linear combination of the parabolic cylinder functions (as the 
modes $v_k$ themselves can be). 
But, we find that the generic solution for~$z$ does not remain positive
definite, which is unacceptable (due to the form of $z$ quoted above).
Therefore, the proposal does not seem viable.  
If we now instead consider a massive, test scalar field of mass~$\mu$ in 
a radiation dominated universe, one arrives at an equation governing the 
modes exactly as in eq.~\eqref{eq:de-i}. 
Interestingly, one indeed obtains a spectrum of particles as in 
eq.~\eqref{eq:beta-se} when the evolution of massive scalar fields are 
examined in certain scenarios involving radiation dominated universes 
(in this context, see ref.~\cite{Audretsch:1978qu}).
If such a scenario is acceptable, there still remains the task of
converting the Gaussian distribution for $\vert\beta(k)\vert^2$ 
into a lognormal distribution.
Remarkably, if we replace $k^2$ by $f^2(k)$ with $f(k)=\mathrm{ln}\,(k/\kf)$,
we indeed arrive at a $\vert\beta(k)\vert^2$ which has a lognormal shape.
However, the challenge is to justify the replacement of~$k^2$ by a generic
function $f^2(k)$.
At first sight this seems possible if we modify the dispersion relation so
that $\omega^2(k)=k^2$ is replaced by $\omega^2(k)=f^2(k)$.
However, note that, since the field is evolving in a FLRW universe, such a 
modified dispersion relation would apply to the physical wave number $k/a$ 
rather than to $k$ itself (in this context, see the discussion in
ref.~\cite{Brandenberger:2012aj}).
Clearly, such a choice modifies the equation~\eqref{eq:de-i} and hence 
the solutions completely.
More importantly, as we pointed out, it has been established that strong
modifications to the dispersion relation will lead to a copious amount of 
particle production which backreacts significantly on the background (in 
this context, also see the following subsection on the issue of 
backreaction).
The above set of arguments suggests that it is rather difficult to 
construct mechanisms that lead to the form of $\beta(k)$ that we have 
worked with.


\subsection{Limits due to backreaction}

In this subsection, we shall discuss another challenge that arises with the 
squeezed initial states we have worked with.
When the perturbations are evolved from non-vacuum initial states, we must 
ensure that the energy density associated with the excited states is 
less than the energy density driving the inflationary background. 
If the densities become comparable, then, evidently, the perturbations can 
start affecting the background dynamics. 
This issue is often referred to as the backreaction problem (see, for instance,
refs.~\cite{Porrati:2004gz,Collins:2006bg,Holman:2007na,Brandenberger:2012aj,
Shukla:2016bnu,Albrecht:2018hoh}).
We shall now arrive at constraints on the parameter~$\gamma$ that determines 
the strength of the squeezed states by demanding that the issue of backreaction
is avoided in the situation we are considering.

The task ahead is to calculate the energy density associated with the 
curvature perturbations when they are assumed to be in a squeezed
initial state.
We find that the energy density associated with the curvature perturbations 
in the de Sitter limit that we are considering can be expressed as follows:
\begin{eqnarray}
\rho_{_{\mathcal{R}}}
&=&\rho_{_{\mathcal{R}}}^{(1)}+\rho_{_{\mathcal{R}}}^{(2)}\nn\\
& \simeq & \f{1}{2\,\pi^2\,a^4}\int_{-\eta^{-1}}^{\infty}\,
\d k\,k^3\,\vert\beta(k)\vert^2\nn\\
& &+\,\frac{\HI^2}{8\,\pi^2\,a^2}\int_{0}^{-\eta^{-1}}\,
dk\,k\,\biggl\{2\,\vert\beta(k)\vert^2-\l[\alpha(k)\,\beta^\ast(k)
+\alpha^\ast(k)\,\beta(k)\r]\biggr\}.\label{eq:rho-p}
\end{eqnarray}
where $\beta(k)$ is the Bogoliubov coefficient which indicates the extent 
of deviation from the Bunch-Davies vacuum.
There are a couple of clarifying remarks we should make regarding this
expression.
Firstly, in arriving at the above expression, we have subtracted the contribution 
due to the Bunch-Davies vacuum, which, upon regularization, is known to correspond 
to (see, for example, refs.~\cite{Allen:1987tz,Anderson:2000wx}) 
\begin{equation}
\rho_{_{\mathcal{R}}}^{_{\mathrm{BD}}}
=\f{61\,\HI^4}{960\,\pi^4}.
\end{equation}
Clearly, this is sub-dominant to the background energy density which behaves
as $\rho_{_{\mathrm{I}}}=3\,\HI^2\,\Mpl^2$ (since $\HI/\Mpl < 10^{-5}$).
Secondly, it should be evident that we have divided the total energy density 
into two parts, with the first part $\rho_{_{\mathcal{R}}}^{(1)}$ arising 
from the contributions due to the modes that are in the sub-Hubble domain 
at any instance, while the second part $\rho_{_{\mathcal{R}}}^{(2)}$ 
corresponds to modes that are in the super-Hubble domain.
At early times, when all the modes are well inside the Hubble radius, it 
is the first part that dominates (in this context, see, for instance,
refs.~\cite{Holman:2007na,Kundu:2013gha}).
This result can be easily understood in simple instances such as, say, power 
law inflation.
In such cases, as is well known, the curvature perturbation behaves in a 
manner similar to that of a massless scalar field.
The expression~$\rho_{_{\mathcal{R}}}^{(1)}$ is essentially the same as the
energy density of a massless scalar field in the sub-Hubble limit.
Note that the energy density~$\rho_{_{\mathcal{R}}}^{(1)}$ 
behaves as $a^{-4}$.
In other words, the energy density is the largest at early times when the 
initial conditions are imposed on the modes of interest in the sub-Hubble
regime.
We shall soon see that this behavior severely restricts the amplitude of
the parameter~$\gamma$.

As we discussed above, it is the sub-Hubble contribution~$\rho_{_{\mathcal{R}}}^{(1)}$
that dominates in the expression~(\ref{eq:rho-p}) for $\rho_{_{\mathcal{R}}}$
at early times.
Recall that, in the scenario we are considering, $\beta(k)$ is determined
by the lognormal function~$g(k)$ [cf. eqs.~\eqref{eq:gk-ln}
and~\eqref{eq:al-be}] that describes the feature in the scalar power spectrum. 
Since $g(k)$ is a Gaussian with the strength $\gamma$ at its maximum 
[cf. eq.~\eqref{eq:gk-ln}], we have $g(k) \lesssim \gamma$ for all~$k$. 
We have always worked with values such that $\gamma \lesssim {\cal O}(1)$. 
Therefore, we can approximate the expression for $\beta(k)$ that is to be 
used in the integral describing~$\rho_{_{\mathcal{R}}}^{(1)}$
[cf. eq.~\eqref{eq:rho-p}] as $\beta(k) \simeq -g(k)/2$.
This simplifies the evaluation of~$\rho_{_{\mathcal{R}}}^{(1)}$, and we obtain 
the energy density of the perturbations in terms of the parameters~$\gamma$, 
$\kf$ and $\Delta_k$ to be
\begin{equation}
\rho_{_{\mathcal{R}}}
\simeq \rho_{_{\mathcal{R}}}^{(1)} 
\simeq \frac{\gamma^2\,\mathrm{e}^{4\,\Delta_k^2}}{16\,
\pi^{5/2}\,\Delta_k}\,\l(\f{\kf}{a}\r)^4.
\end{equation}
We should stress again that we have subtracted the contribution due to the 
Bunch-Davies vacuum in arriving at this expression.
Due to this reason, we should also add that no regularization is required 
to arrive at the above result.
Hence, $\rho_{_{\mathcal{R}}}\to 0$ when $\gamma\to 0$, as expected.
We find that the relative difference between the above approximate estimate 
of $\rho_{_{\cal R}}^{(1)}$ [obtained by assuming that $\beta(k) \simeq -g(k)/2$]
and the exact estimate is at most of $\mathcal{O}(1)$. 
Therefore, for convenience, we shall use the approximate estimate to arrive 
at the bound on the parameter $\gamma$ in our scenario.

For the backreaction to be negligible in our scenario, we require that 
$\rho_{_{\mathcal{R}}} \ll \rho_{_\mathrm{I}}$, where, as we mentioned, 
$\rho_{_{\mathrm{I}}} = 3\,\HI^2\,\Mpl^2$ is the energy density of the 
background during inflation. 
This requirement leads to the condition
\begin{equation}
\f{\gamma^2\,\mathrm{e}^{4\,\Delta_k^2}}{\Delta_k}\,
\l(\f{\kf}{a\,\HI}\r)^4 \ll 48\,\pi^{5/2}\,\l(\f{\Mpl}{\HI}\r)^2.
\end{equation}
During inflation, the value of the Hubble parameter $\HI$ is related to the 
tensor-to-scalar ratio~$r$ through the relation $(\HI/\Mpl)^2 \simeq r\,
A_{_{\mathrm{S}}}$, where $A_{_{\mathrm{S}}}\simeq 2.11\times10^{-9}$ is the 
COBE normalized scalar amplitude over the CMB scales.
Since the energy density $\rho_{_{\mathcal{R}}}$ is the largest at early times,
let us evaluate it at the time when the smallest wave number of interest, say,
$k_\mathrm{min}$, leaves the Hubble radius, i.e. when 
$k_\mathrm{min}=a_\mathrm{min}\,\HI$.
At such a time, as we have set $\Delta_k=1$, the above inequality reduces to 
(upon ignoring the constant coefficients)
\begin{equation}
\gamma\ll \f{10^{9/2}}{\sqrt{r}}\,\, \l(\f{k_\mathrm{min}}{\kf}\r)^2.
\end{equation}
It seems reasonable to set $k_\mathrm{min}=k_1/10 \simeq 10^{-5}\,\mpcinv$ 
(recall that we had earlier chosen $k_1=10^{-4}\,\mpcinv$).
If we choose $\kf=10^{5}\, \mpcinv$, which is the smallest of the values for~$\kf$
that we had considered, then we arrive at $\gamma\ll 10^{-16.5}/\sqrt{r}$.
In other words, for $r\simeq 10^{-3}$, we require $\gamma < 10^{-15}$.
For a larger $\kf$, clearly, the limits on $\gamma$ are even stronger.
If $\kf \simeq 10^{13}\, \mpcinv$ and  $r\simeq 10^{-3}$, we require
that $\gamma < 10^{-30}$.
Evidently, $\gamma$ can be larger if the tensor-to-scalar ratio is smaller, i.e.
when the scale of inflation is lower.
Nevertheless, even for an extreme value of $r \simeq 10^{-30}$ as suggested
by the recent arguments based on the trans-Planckian censorship conjecture (in 
this context, see, for instance, ref.~\cite{Bedroya:2019tba}), we require $\gamma 
< 10^{-2}$ for $\kf \simeq 10^5\,\mpcinv$ and $\gamma < 10^{-17}$ for $\kf \simeq
10^{13}\,\mpcinv$.
We have instead worked with $\gamma\simeq 1$ for $\kf=10^{5}\, \mpcinv$ and 
$\gamma\simeq 10^{-8}$ for $\kf=10^{13}\, \mpcinv$.
Clearly, for a more reasonable $r$, the constraints on $\gamma$ are considerably
more severe.
Under such conditions, $\fnl$ and hence the non-Gaussian modifications will prove 
to be small and we will not be able to achieve the desired level of amplification 
of the corrected power spectrum $\ps(k)+\pc(k)$.
In fact, $\gamma$ is so tightly constrained by the backreaction that we are 
essentially left with the slow roll results.

There are two related points we wish to make here.
Firstly, one may wonder if the energy associated with the curvature 
perturbation~$\rho_{_{\mathcal{R}}}$ itself may support accelerated 
expansion.
Since~$\rho_{_{\mathcal{R}}}^{(1)}\propto a^{-4}$, conservation of energy 
suggests that, at early times, the pressure associated with the excited states 
should be given by $p_{_{\mathcal{R}}}^{(1)}=\rho_{_{\mathcal{R}}}^{(1)}/3$.
Upon explicit calculation, we find that this is indeed the case (in 
this context, also see refs.~\cite{Kundu:2013gha,Shukla:2016bnu}).
In other words, the pressure associated with the excited initial states 
does not possess the equation of state required to drive inflation.
Secondly, since the energy density of the perturbations~$\rho_{_{\mathcal{R}}}^{(1)}$
dies down as $a^{-4}$, one may imagine that it could decay rapidly enough 
permitting the background energy density to dominate.
Given~$\rho_{_{\mathcal{R}}}$ at $a=a_\mathrm{min}$, we find that the 
number of e-folds after which the energy density associated with the 
perturbations becomes sub-dominant to $\rho_{_{\mathrm{I}}}$ is given by
\begin{equation}
N\simeq \f{1}{4}\,\mathrm{ln}\l(\f{\gamma^2\, r}{10^9}\r)
+\mathrm{ln}\l(\f{\kf}{k_{\mathrm{min}}}\r).
\end{equation}
For the values of the various quantities we have worked with, say, $\gamma
\simeq 1$, $\kf=10^{5}\,\mpcinv$ and $k_{\mathrm{min}}=10^{-5}\,\mpcinv$, 
if we choose a tensor-to-scalar ratio of $r\simeq 10^{-3}$, we find that it 
will take as many as $16$ e-folds before the background energy density 
begins to dominate.
This duration will be more prolonged for larger values of~$\kf$.
Clearly, backreaction is a rather serious issue that needs to be accounted 
for.


\section{Conclusions}\label{sec:f}

In this work, we had explored a possible mechanism for the production 
of PBHs and GWs wherein the primordial scalar perturbations were evolved 
from squeezed initial states. 
The advantage of the mechanism is the fact that it is completely
independent of the actual model that drives the background dynamics 
during inflation. 
All we require is typical slow roll inflation which leads to a power 
spectrum that is consistent with the recent CMB data on large scales. 
By choosing specific forms for the Bogoliubov coefficients that characterize
the squeezed states, we had constructed scalar power spectra with a lognormal 
feature at small scales.
It is well known that, in such cases, the scalar bispectra in the squeezed 
limit is inversely proportional to the value of the squeezed mode, a dependence
which we expected to utilize so that we obtain significantly high values for 
the scalar non-Gaussianity parameter~$\fnl$ at large wave numbers.
We had hoped that this property can lead to large non-Gaussian modifications to 
the scalar power spectrum, which in turn can amplify the power considerably at
small scales. 
While the proposal seemed feasible, there were two challenges that we had encountered.
Mathematically, it was rather easy to construct squeezed initial states that led 
to a sharp rise in power on small scales, when the non-Gaussian modifications were 
taken into account.
However, we had found that it can be a challenge to design scenarios that excite 
the curvature perturbation to such an initial state during the early stages of 
inflation.
Moreover, we had found that the backreaction on the inflationary background due 
to the excited state of the perturbations strongly limits the extent of deviation 
from the Bunch-Davies vacuum.
In fact, the bounds due to the backreaction are so strong that the slow roll 
results remain valid.

Let us make a few further clarifying remarks at this stage of our discussion.
The consistency condition relating the bispectrum and the power spectrum is 
known to be violated for modes that evolve from the non-vacuum initial states 
(i.e. around the peaks in the original power spectra).
As a result, we had expected that the contributions to the non-Gaussianity 
parameter due to the so-called local observer effect that has to be 
subtracted will be small when compared to the actual value $\fnl$ over these 
wave numbers (in this context, see refs.~\cite{Tada:2016pmk,Suyama:2020akr}).
Motivated by the largely local form of the scalar bispectrum in the squeezed
limit, we had utilized the corresponding $\fnl$ to calculate the non-Gaussian
modifications to the power spectrum~\cite{Motohashi:2017kbs,Cai:2018dig,
Unal:2018yaa,Ragavendra:2020sop}.
We had hoped that the non-Gaussian modifications will dominate leading to 
enhanced power at small scales.
However, we had found that the issue of backreaction put paid to the proposal.

Before we conclude, we would like to comment on four issues and their possible 
resolutions in the approach of generating PBHs and GWs from squeezed initial 
states.
\begin{enumerate}
\item 
Note that we have arrived at the scalar bispectrum by calculating the 
integrals involved over the domain $-\infty < \eta <0$.
In other words, we have assumed that the initial squeezed state was
chosen in the infinite past, i.e. as $\eta \to -\infty$.
It may be argued that if we choose to work with non-vacuum initial
states, then the initial conditions need to be imposed at a finite
initial time, say, $\eta_\mathrm{i}$.
We believe that our results and conclusions will hold as long as 
$\eta_\mathrm{i}\ll -1/k_\mathrm{min}$, where, recall that, we have set
$k_\mathrm{min} \simeq k_1/10$, with $k_1$ being the smallest wave number 
of observational interest, which we have assumed to be $10^{-4}\,\mpcinv$.
\item
The method by which we have calculated modifications to the power spectrum 
due to the scalar non-Gaussianity parameter is strictly valid for an $\fnl$ 
of the local type.
In other words, $\fnl$ ought to be a constant independent of scale. 
However, in our scenario, the $\fnl$ we obtain is strongly scale dependent.
There are two points that we believe support the method we have adopted. 
Firstly, in order to mimic the local behavior of $\fnl$, we have chosen 
to work with its value in the squeezed limit (in this context, also see
ref.~\cite{Motohashi:2017kbs}).
Secondly, and interestingly, we find that, near the wave numbers corresponding
to the peaks of the power spectra, the non-Gaussianity parameter $\fnl$ seems
to have a strongly local shape.
We should add here that a formal approach to arrive at the modifications to the 
power spectrum would be to calculate the loop corrections at the appropriate order.
While such an effort seems worthwhile, we believe that, since the parameter~$\fnl$ 
is largely local around the maximum in the power spectrum, our calculations 
can be considered to be fairly suggestive.
\item
In our approach, we have accounted for the cubic order
non-Gaussianities by considering the corresponding modifications to the 
scalar power spectrum.
This approach seems adequate to account for the non-Gaussian modifications 
to the density parameter~$\ogw$ describing the stochastic GW 
background~\cite{Garcia-Bellido:2017aan,Cai:2018dig,Unal:2018yaa}.
However, when calculating the density of PBHs formed, the non-Gaussianities 
are expected to also modify the probability distribution of the density 
contrast and hence the number of PBHs at the time of their formation 
[cf. eq.~\eqref{eq:b-pbh}].
We should mention that this effect needs to be accounted for 
separately~\cite{Germani:2019zez}.
\item 
Lastly, it may be interesting to explore if the contributions due to the higher 
order correlations such as the trispectrum may rescue our proposal and lead to 
large non-Gaussian modifications despite the strong constraints on $\gamma$ due 
to the backreaction~\cite{Nakama:2016gzw,Yuan:2020iwf}.
For instance, we had seen that, in the squeezed limit, $\fnl$ had behaved 
as~$\kf/k_1$.
If the non-Gaussianity parameter, say, $\tau_{_{\mathrm{NL}}}$, characterizing
the trispectrum (in this context, see ref.~\cite{Seery:2006js}) in a squeezed
initial state behaves in a stronger fashion, it seems possible that the higher
order terms may modify the power spectrum adequately to circumvent the limits 
on~$\gamma$.
However, even if this works out, one concern would remain.
We had seen that, despite the large value of $\fnl$, the amplitude of the 
modified power spectrum was of the order of $10^{-2}$ (for the original
values of $\gamma$ we had worked with).
If the non-Gaussian modifications due to the trispectrum prove to be significant,
it is possible that these higher order contributions will also affect the validity 
of perturbation theory. 
One will have to ensure that the amplitude of the corrected power spectrum remains 
smaller than unity even when further contributions are taken into account.
Probably, the conditions for the validity of the perturbation theory at higher orders 
would severely restrict the extent of deviations from the Bunch-Davies vacuum.
We are currently exploring these issues.
\end{enumerate}
We would like to close by pointing out that, the various arguments
we have considered in this work suggest that the initial state of the curvature 
perturbations is likely to be remarkably close to the Bunch-Davies vacuum, 
in particular, on small scales.


\acknowledgments

HVR and LS wish to thank Debika Chowdhury and V.~Sreenath for discussions.
The authors wish to thank Ilia Musco for comments on the manuscript.
HVR would also like to thank the Indian Institute of Technology Madras, 
Chennai, India, for support through the Half-Time Research Assistantship.
LS wishes to acknowledge support from the Science and Engineering 
Research Board, Department of Science and Technology, Government of 
India, through the Core Research Grant CRG/2018/002200.


\appendix


\section{The dominant contributions to the scalar bispectrum}\label{app:G-ab}

In this appendix, we shall provide the complete expressions describing the 
dominant contributions to the scalar bispectrum evaluated in a squeezed
initial state. 
For a generic $\alpha(k)$ and $\beta(k)$, these contributions are given by   
the following expressions (in this context, see for example, 
refs.~\cite{Meerburg:2009ys,Ganc:2011dy,Brandenberger:2012aj,Kundu:2013gha}):
\begin{subequations}\label{eq:G-ce}
\begin{eqnarray}
G_1(\vka,\vkb,\vkc) 
&=& \frac{\HI^4}{32\,\Mpl^4\,\epsilon_1}\,
\frac{\vert\alpha_1\vert^2\,\vert\alpha_2\vert^2\,
\vert\alpha_3\vert^2}{k_1\,k_2\,k_3}
\frac{(1 - \delta_1)\,(1 - \delta_2)\,(1 - \delta_3)}{k_1^2}\,\nn\\
& &\times\, \Biggl[\frac{1+\delta^\ast_1\,\delta^\ast_2\,\delta^\ast_3}{\kT}\,
\l(1+ \frac{k_1}{\kT}\r)
+ \frac{\delta^\ast_1 + \delta^\ast_2\,\delta^\ast_3}{k_1 - k_2 - k_3}\,
\l(1+\f{k_1}{k_1 - k_2 - k_3}\r)\nn\\
& &+\, \frac{\delta^\ast_2 +\delta^\ast_1\,\delta^\ast_3}{k_2 - k_1 - k_3}
\l(1-\f{k_1}{k_2 - k_1 - k_3}\r)
+ \frac{\delta^\ast_3 +\delta^\ast_1\,\delta^\ast_2}{k_3 - k_1 - k_2}
\l(1-\f{k_1}{k_3 - k_1 - k_2}\r)\Biggr]\nn\\
& &+\, \mathrm{complex~conjugate}+ \mathrm{two~permutations},\label{eq:G1_ab}\\
G_2(\vka,\vkb,\vkc)
&=& -\f{\HI^4}{64\,\Mpl^4\,\epsilon_1}\,
\f{(k_1^2 + k_2^2 + k_3^2)}{(k_1\,k_2\,k_3)^3}\,
\vert\alpha_1\vert^2\,\vert\alpha_2\vert^2\,\vert\alpha_3\vert^2\,
\l(1 - \delta_1\r)\,\l(1 - \delta_2\r)\,\l(1 - \delta_3\r)\nn\\
& &\times\,\Biggl\{\Bigg[\f{i}{\ee}\,\Biggl(\mathrm{e}^{i\,\kT\,\ee} 
-\delta^\ast_1\,\mathrm{e}^{i\,(-k_1+k_2 + k_3)\,\ee} 
- \delta^\ast_2\,\mathrm{e}^{i\,(k_1 - k_2 + k_3)\,\ee} 
- \delta^\ast_3\,\mathrm{e}^{i\,(k_1 + k_2 - k_3)\,\ee}\nn\\
& &+\, \delta^\ast_2\,\delta^\ast_3\,\mathrm{e}^{-i\,(-k_1+k_2 + k_3)\,\ee}
+ \delta^\ast_1\,\delta^\ast_3\,\mathrm{e}^{-i\,(k_1 - k_2 + k_3)\,\ee} 
+ \delta^\ast_1\,\delta^\ast_2\,\mathrm{e}^{-i\,(k_1 + k_2 - k_3)\,\ee}\nn\\
& &-\, \delta^\ast_1\,\delta^\ast_2\,\delta^\ast_3\,
\mathrm{e}^{-i\,\kT\,\ee}\Biggr)\Biggr]_{\ee\to 0}
+\, \l(1 + \delta^\ast_1\,\delta^\ast_2\,\delta^\ast_3\r)\,
\f{(k_1\,k_2 + k_2\,k_3 + k_1\,k_3)}{\kT}\nn\\
& &+\, \l(\delta^\ast_1 + \delta^\ast_2\,\delta^\ast_3\r)\,
\f{(k_1\,k_2 - k_2\,k_3 + k_1\,k_3)}{(-k_1+k_2 + k_3)}\nn\\
& &+\, \l(\delta^\ast_2 + \delta^\ast_1\,\delta^\ast_3\r)\,
\f{(k_1\,k_2 + k_2\,k_3 - k_1\,k_3)}{(k_1 - k_2 + k_3)}\nn\\
& &+\, \l(\delta^\ast_3 + \delta^\ast_1\,\delta^\ast_2\r)\,
\f{(-k_1\,k_2+k_2\,k_3 + k_3\,k_1)}{(k_1 + k_2 - k_3)}\nn\\
& &+\,k_1\,k_2\,k_3\,
\Biggl[\f{1 +\delta^\ast_1\,\delta^\ast_2\,\delta^\ast_3}{\kT^2} 
+ \f{(\delta^\ast_1 + \delta^\ast_2\,\delta^\ast_3)}{(-k_1 + k_2 + k_3)^2}
+ \f{(\delta^\ast_2 + \delta^\ast_1\,\delta^\ast_3)}{(k_1 - k_2 + k_3)^2}\nn\\ 
& &+\, \f{(\delta^\ast_3 + \delta^\ast_1\delta^\ast_2)}{(k_1 + k_2 - k_3)^2}\Biggr]
+ \mathrm{complex~conjugate},\label{eq:G2_ab}\\
G_3(\vka,\vkb,\vkc) 
&=&-\f{\HI^4}{32\,\Mpl^4\,\epsilon_1}\,
\f{\vert\alpha_1\vert^2\,\vert\alpha_2\,\vert^2\,\vert\alpha_3\vert^2}{k_1\,
k_2\,k_3}\,
\frac{(1 - \delta_1)\,(1 - \delta_2)\,(1 - \delta_3)}{k_1^2}\nn\\
& &\times\,\f{(k_2^2 - k_3^2)^2 - k_1^2\,(k_2^2 + k_3^2)}{2\,k_2^2\,k_3^2} \nn\\
& &\times\, \Biggl[\frac{1+\delta^\ast_1\,\delta^\ast_2\,\delta^\ast_3}{\kT}\,
\l(1+ \frac{k_1}{\kT}\r)
+ \frac{\delta^\ast_1 + \delta^\ast_2\,\delta^\ast_3}{k_1 - k_2 - k_3}\,
\l(1+\f{k_1}{k_1 - k_2 - k_3}\r)\nn\\
& &+\, \frac{\delta^\ast_2 +\delta^\ast_1\,\delta^\ast_3}{k_2 - k_1 - k_3}
\l(1-\f{k_1}{k_2 - k_1 - k_3}\r)
+ \frac{\delta^\ast_3 +\delta^\ast_1\,\delta^\ast_2}{k_3 - k_1 - k_2}
\l(1-\f{k_1}{k_3 - k_1 - k_2}\r)\Biggr]\nn\\
& &+\, \mathrm{complex~conjugate}+ \mathrm{two~permutations},\label{eq:G3_ab}\\
G_7(\vka,\vkb,\vkc) 
&=& \frac{\HI^4\,\epsilon_2}{32\,\Mpl^4\epsilon_1^2}\,
\biggl\{\frac{1}{(k_1\,k_2)^3}\,\vert\alpha_1\vert^2\,\vert\alpha_2\vert^2\,
\l(1 - \delta_1\r)\,\l(1 - \delta^\ast_1\r)\,
\l(1 - \delta_2\r)\,\l(1 - \delta^\ast_2\r)\nn\\
& &+\, {\rm two~permutations}\biggr\},\label{eq:G7_ab}
\end{eqnarray}
\end{subequations}
where $\kT =k_1+k_2+k_3$ and, for convenience, we have set $\alpha(k_i)=\alpha_i$ 
and $\delta(k_i)=\delta_i$ for $i=\{1,2,3\}$.
Note that, we can write 
\begin{equation}
\vert\alpha(k)\vert^2=\l[1-\vert\delta(k)\vert^2\r]^{-1}
\end{equation}
so that the complete bispectrum can be expressed in terms of the 
function~$\delta(k)$,
which in turn is determined by the feature $g(k)$ in the power spectrum
[cf. eqs.~\eqref{eq:gk} and~\eqref{eq:deltak}].
\bibliographystyle{JHEP}
\bibliography{pbh-sgw-ab}
\end{document}